\documentclass[fleqn,usenatbib]{mnras}

\usepackage[T1]{fontenc}

\DeclareRobustCommand{\VAN}[3]{#2}
\let\VANthebibliography\thebibliography
\def\thebibliography{\DeclareRobustCommand{\VAN}[3]{##3}\VANthebibliography}

\usepackage{graphicx}	
\usepackage{amsmath}	
\usepackage{amssymb}	

\usepackage{aas_macros}

\title[A sample of localised Fast Radio Bursts]{A sample of Fast Radio Bursts discovered and localised with MeerTRAP at the MeerKAT telescope}

\author[F.~Jankowski et al.]{F.~Jankowski,$^{1,2}$\thanks{E-mail: fabian.jankowski@cnrs-orleans.fr}
M.~C.~Bezuidenhout,$^{1,3}$
M.~Caleb,$^{1,4,5}$
L.~N.~Driessen,$^{1,6}$
M.~Malenta,$^{1}$
V.~Morello,$^{1}$
\newauthor
K.~M.~Rajwade,$^{1,7}$
S.~Sanidas,$^{1}$
B.~W.~Stappers,$^{1}$
M.~P.~Surnis,$^{1,8}$
E.~D.~Barr,$^{9}$
W.~Chen,$^{9}$
\newauthor
M.~Kramer,$^{9,1}$
J.~Wu,$^{9}$
S.~Buchner,$^{10}$
M.~Serylak,$^{11}$
and J.~Xavier~Prochaska$^{12,13}$
\\
$^{1}$Jodrell Bank Centre for Astrophysics, Department of Physics and Astronomy, The University of Manchester, Manchester M13 9PL, UK\\
$^{2}$LPC2E, Universit\'{e} d'Orl\'{e}ans, CNRS, 3A Avenue de la Recherche Scientifique, 45071 Orl\'{e}ans, France\\
$^{3}$Centre for Space Research, North-West University, Potchefstroom 2531, South Africa\\ 
$^{4}$Sydney Institute for Astronomy, School of Physics, The University of Sydney, NSW 2006, Australia\\
$^{5}$ASTRO3D: ARC Centre of Excellence for All-sky Astrophysics in 3D, Canberra 2601, ACT, Australia\\
$^{6}$CSIRO, Space and Astronomy, PO Box 1130, Bentley, WA 6102, Australia\\
$^{7}$ASTRON, the Netherlands Institute for Radio Astronomy, Oude Hoogeveensedijk 4, 7991 PD Dwingeloo, The Netherlands\\
$^{8}$Department of Physics, IISER Bhopal, Bhauri Bypass Road, Bhopal, 462066, India\\
$^{9}$Max-Planck-Institut f{\"u}r Radioastronomie, Auf dem H{\"u}gel 69, D-53121 Bonn, Germany\\
$^{10}$South African Radio Astronomy Observatory, Black River Park, 2 Fir Street, Observatory, Cape Town, 7925, South Africa\\
$^{11}$The Square Kilometre Array Observatory, Lower Withington, Macclesfield, Cheshire, SK11 9FT, UK\\
$^{12}$University of California, Santa Cruz, 1156 High Street, Santa Cruz, CA 95064, USA\\
$^{13}$Kavli Institute for the Physics and Mathematics of the Universe, 5-1-5 Kashiwanoha, Kashiwa, 277-8583, Japan\\
}

\date{Accepted XXX. Received YYY; in original form ZZZ}

\pubyear{2023}

\begin{document}
\label{firstpage}
\pagerange{\pageref{firstpage}--\pageref{lastpage}}
\maketitle


\begin{abstract}
We present a sample of well-localised Fast Radio Bursts (FRBs) discovered by the MeerTRAP project at the MeerKAT telescope in South Africa. We discovered the three FRBs in single coherent tied-array beams and localised them to an area of $\sim$1~$\text{arcmin}^2$. We investigate their burst properties, scattering, repetition rates, and localisations in a multi-wavelength context. FRB~20201211A shows hints of scatter broadening but is otherwise consistent with instrumental dispersion smearing. For FRB~20210202D, we discovered a faint post-cursor burst separated by $\sim$200~ms, suggesting a distinct burst component or a repeat pulse. We attempt to associate the FRBs with host galaxy candidates. For FRB~20210408H, we tentatively (0.35 - 0.53 probability) identify a compatible host at a redshift $\sim$0.5. Additionally, we analyse the MeerTRAP survey properties, such as the survey coverage, fluence completeness, and their implications for the FRB population. Based on the entire sample of 11 MeerTRAP FRBs discovered by the end of 2021, we estimate the FRB all-sky rates and their scaling with the fluence threshold. The inferred FRB all-sky rates at 1.28~GHz are $8.2_{-4.6}^{+8.0}$ and $2.1_{-1.1}^{+1.8} \times 10^3 \: \text{sky}^{-1} \: \text{d}^{-1}$ above 0.66 and 3.44~Jy~ms for the coherent and incoherent surveys, respectively. The scaling between the MeerTRAP rates is flatter than at higher fluences at the 1.4-$\sigma$ level. There seems to be a deficit of low-fluence FRBs, suggesting a break or turn-over in the rate versus fluence relation below 2~Jy~ms. We speculate on cosmological or progenitor-intrinsic origins. The cumulative source counts within our surveys appear consistent with the Euclidean scaling.
\end{abstract}

\begin{keywords}
transients: fast radio bursts -- surveys -- methods: data analysis -- radiation mechanisms: non-thermal -- techniques: interferometric
\end{keywords}



\section{Introduction}
\label{sec:introduction}

Fast Radio Bursts (FRBs) are extremely luminous, approximately millisecond-duration bursts of radio emission originating from cosmological distances at inferred redshifts of up to a few. First discovered in 2007 \citep{2007Lorimer} and confirmed as a population by \citet{2013Thornton}, there are now more than 600 FRBs published on the Transient Name Server\footnote{\url{https://www.wis-tns.org/}}. Despite this significant increase in sample size, primarily driven by surveys with wide-field radio interferometers such as the Canadian Hydrogen Intensity Mapping Experiment (CHIME; \citealt{2021CHIMECatalogue}), we still do not know what physical mechanism creates FRBs. This and their unknown origins are currently one of the most interesting topics in radio astronomy and astrophysics \citep{2019Petroff, 2022Petroff}. While the vast majority of FRBs seem to be one-off bursts that could have resulted from cataclysmic events like compact-object mergers or (stellar) explosions \citep{2019Platts}, the discovery of repeating FRBs \citep{2016Spitler} suggested a non-cataclysmic origin for at least some of them. The discovery in 2020 of FRB-like bursts, some contemporary with hard X-ray emission, from the Galactic magnetar SGR~J1935+2154 \citep{2020Bochenek, 2020CHIMESGR, 2021LiSGR} established a connection between at least some repeating FRBs and magnetars. Unfortunately, the transient nature of FRBs makes them hard to study and requires enormous amounts of observing time that can only realistically be afforded through commensal surveys. Additionally, there are currently only 24 repeating FRBs published (about 4~per cent of the current FRB population) with reasonably precise on-sky localisations that allow us to study the FRB emission process in detail using dedicated multi-frequency observations. While the repeaters provide great opportunities for targeted and long-term follow-up, the sample is possibly biased and not representative of the whole population. Additionally, the repeating FRBs may form a separate FRB class altogether \citep{2021CHIMECatalogue}. Consequently, most of the research endeavour still lies in discovering and characterising new one-off FRBs and expanding the sample of repeating sources. Survey projects at various radio telescopes have driven the field in the last few years, for instance, SUPERB \citep{2018Keane} at the Parkes \textit{Murriyang} radio telescope, the UTMOST FRB search project \citep{2017Bailes} at the Molonglo Synthesis Radio Telescope (MOST), the Commensal Real-time ASKAP Fast Transients (CRAFT; \citealt{2010Macquart}) survey at the Australian Square Kilometre Array Pathfinder (ASKAP), the CHIME/FRB project \citep{2018CHIMEFRBsurvey} at CHIME, the Apertif Radio Transient System (ARTS; \citealt{2014VanLeeuwen}) project at the Westerbork Synthesis Radio Telescope (WSRT), and the Commensal Radio Astronomy FAST Survey (CRAFTS; \citealt{2018Li}) at the Five-hundred-meter Aperture Spherical Telescope (FAST). Several other facilities are currently in the design, commissioning, or early-science phase, such as the Deep Synoptic Array (DSA; \citealt{2019Kocz}) or the Canadian Hydrogen Observatory and Radio-transient Detector (CHORD; \citealt{2019Vanderlinde}).

While the total number of published FRBs is already sufficient to enable the first meaningful population studies \citep{2021CHIMECatalogue}, the vast majority of them are too poorly localised for deep optical imaging or follow-up observations with sensitive narrow field-of-view (FoV) instruments in other wavebands. More importantly, their poor radio localisations prevent us from robustly associating them with their host galaxies and thereby measuring their redshifts \citep{2017Eftekhari}. Robust FRB to host associations are usually characterised by low chance coincidence probabilities $< 0.1$ \citep{2017Eftekhari, 2020Heintz}, or conversely, high association probabilities $> 0.95$ \citep{2021Aggarwal}. On the other hand, the CRAFT team has been increasingly successful at localising one-off bursts to their host galaxies using ASKAP \citep{2019Bannister, 2020Macquart}. Similarly, several repeaters have been localised to milliarcsecond precision using multi-station very-long baseline interferometry (VLBI) \citep{2017Chatterjee, 2020Marcote}, for example, as part of the PRECISE project (Pinpointing REpeating ChIme Sources with EVN dishes; \citealt{2022Marcote}). From the above, it is clear that higher precision radio interferometric localisations are needed to advance the field.

The fact that FRBs are bright and temporally narrow radio pulses makes them excellent probes of the intervening ionised media \citep{2020Macquart}. The turbulent plasmas that an FRB traverses from its host to the observer imprint characteristic signatures onto its radio signal through propagation effects such as dispersion, pulse broadening (scattering), scintillation, refraction (lensing), or absorption \citep{2019Cordes}. While its dispersion measure (DM) is a proxy for distance assuming various Galactic and extragalactic free-electron models \citep{2002Cordes, 2017Yao, 2020Yamasaki, 2018Zhang, 2020Macquart}, the observed pulse broadening encodes the turbulence, distribution, and scattering geometries of intervening plasmas. Measured scattering times allow us to estimate host galaxy DM contributions, act as a combined DM -- scattering time estimator for host galaxy redshifts, or can constrain the intergalactic medium's (IGM) baryonic fraction if the host redshift is known \citep{2022Cordes}. FRB scatter broadening is, therefore, an important measurable quantity. Unfortunately, it is often challenging to measure or disentangle various contributions, as the observed FRB signal is the convolution product of the emitted burst with several astrophysical line-of-sight and instrumental components.

In this paper, we present a sample of well-localised FRBs discovered in the commensal MeerTRAP transient survey running at the 64-element MeerKAT telescope array in South Africa. The formation of hundreds of coherent tied-array beams inside the MeerKAT primary beam allowed us to localise them to about $1~\text{arcmin}^2$ or better. These are more precisely localised than the vast majority of FRBs currently published.

In the following, we describe the MeerTRAP transient surveys and the data presented in \S~\ref{sec:observations}. In \S~\ref{sec:analysis}, we discuss the techniques employed in our burst analysis, FRB localisation, host galaxy association, and survey characterisation. In \S~\ref{sec:results}, we present the FRB sample discovered, their burst properties, localisations within a multi-wavelength context, and our inferences from the MeerTRAP surveys, such as the FRB all-sky rate and its scaling with burst fluence. In \S~\ref{sec:discussion}, we discuss our results compared with the literature. Finally, we summarise our results and present our conclusions in \S~\ref{sec:conclusions}. Throughout the paper we quote uncertainties at the 1-$\sigma$ level if not stated otherwise, employ the parameters of the ``Planck 2018'' cosmology \citep{2020Planck}, and use an inverse dispersion constant rounded to three significant figures ($1/D = 2.41 \times 10^{-4}~\text{MHz}^{-2}~\text{pc}~\text{cm}^{-3}~\text{s}^{-1}$), as is conventional in pulsar astronomy.

\section{The MeerTRAP surveys at the MeerKAT telescope}
\label{sec:observations}

The FRBs presented here are from the recently-commissioned MeerTRAP (More Transients and Pulsars) instrument at the MeerKAT (More Karoo Array Telescope) array in South Africa. They were discovered in the fully-commensal MeerTRAP survey that piggybacks all MeerKAT Large Survey Projects (LSPs) and some other smaller proposals. The FRBs were found in a short period between 2020 December and 2021 April. The MeerKAT telescope is a state-of-the-art interferometric array of 64 dishes of 13.96-m diameter each that are located in the Karoo desert area in South Africa and are operated by the South African Radio Astronomy Observatory (SARAO) \citep{2016Jonas, 2020Mauch}. It is a direct precursor to the mid-frequency component of the Square Kilometre Array. Details of the MeerTRAP system have previously been reported in \citet{2018Sanidas}, \citet{2020Malenta}, \citet{2020Rajwade}, and \citet{2022Jankowski}. A full system overview will be presented in an upcoming publication (Stappers et al.\ in prep.). The discoveries of three other MeerTRAP FRBs were reported in earlier work \citep{2022Rajwade}. Another output from the MeerTRAP project is the discovery of several dozens of Galactic sources, such as canonical radio pulsars and Rotating Radio Transients (RRATs) (\citealt{2022Bezuidenhout}; MeerTRAP in prep.), and an ultra-long period neutron star with a spin period of 76~s \citep{2022Caleb}.

We summarise the most important aspects of the data and data processing system relevant to this work. While MeerTRAP has been involved in observations at both centre frequencies currently supported by MeerKAT, i.e.\ UHF (544 -- 1088 MHz) and L-band (856 -- 1712 MHz), the FRBs presented here were all discovered at L-band frequencies. In particular, the data were obtained in a band of 856~MHz centred at 1284~MHz with a maximum of $\sim$770 MHz on-sky bandwidth. They have a sampling time of 306.24~$\mu \text{s}$, 1024 frequency channels, a channel bandwidth of $\sim$0.836~MHz, and represent total intensity, i.e.\ Stokes I. Two MeerTRAP FRB surveys are running simultaneously. The first one uses the wider FoV but less sensitive MeerKAT primary or incoherent beam (IB) that results from the incoherent combination of the data streams from all the available MeerKAT antennas included in the sub-array that MeerTRAP was commensal with. The central region of the IB is typically tesselated with 768 ($64 \times 12$) tied-array coherent beams (CBs) that are created by beam-forming the voltage data streams from the individual telescopes, i.e.\ the coherent addition of their signals by a dedicated beam-forming instrument known as FBFUSE (Filterbanking Beamformer User Supplied Equipment; \citealt{2018Barr, 2021Chen}). The CBs are arranged in a hexagonal pattern starting from the centre of the IB, meaning that they are close to the maximum sensitivity area of the primary beam. Usually, the CBs are formed from the 40 innermost dishes of the MeerKAT array with a maximum baseline of approximately 800~m. We currently overlap them at 25 per cent of the CB response to increase the sky area tiled with CBs and thereby the FRB yield. Due to the different numbers of antennas contributing in each case, the CBs are approximately $40/\sqrt{64} = 5$ times more sensitive than the IB. We employed the highly-optimized Graphics Processing Unit (GPU)-based \textsc{astroaccelerate} software \citep{2012Armour, 2019Carels, 2020Adamek} to search for dispersed signals in the data stream up to a maximum trial DM of 5118.4~pc~$\text{cm}^{-3}$ and typically up to $\sim$670~ms in boxcar pulse width. We initially considered candidates of all widths but more recently restricted ourselves to candidates up to $\sim$300~ms wide. Before the data were fed to the single-pulse search engine, we automatically excised radio frequency interference (RFI) using a dynamically-changing frequency channel mask. The channel mask was established from the current data block based on how significantly the channels deviated from the median bandpass. This was done using a newly-developed \textsc{iqrm} algorithm and software implementation \citep{2022Morello}. Compared with the static frequency channel masks used in previous work \citep{2022Rajwade}, the fraction of masked channels decreased significantly from typically 50 - 60 to about 20 - 25 per cent. Additionally, we employed a zero-DM filter \citep{2009Eatough} as before. We further processed all single-pulse candidates with signal-to-noise ratios (S/N) $\geq 8.0$. The candidates were clustered in time and DM and were automatically matched with known sources from the literature, such as pulsars and RRATs, using a custom \textsc{python}-based software\footnote{\url{https://github.com/fjankowsk/meertrapdb/}}. We then employed a bespoke image-based machine learning (ML) classifier named \textsc{frbid}\footnote{\url{https://github.com/Zafiirah13/FRBID/}} to classify the candidates into astrophysical pulses and RFI, based on a combination of their trial DM versus time (``bow tie'') images and dedispersed dynamic spectra. \textsc{frbid} was inspired by the \textsc{fetch} transient classifier \citep{2020Agarwal}, but we tuned its features and performance to MeerTRAP data and the particular RFI environment at MeerKAT. We trained \textsc{frbid} on a balanced data set of pulsar and RRAT pulses, the first FRBs, and a selection of RFI recorded by the MeerTRAP backend that our team had visually inspected and manually assigned labels. The training set consisted of about 16,000 candidates, split approximately evenly into genuine transients (FRBs, pulsar and RRAT pulses) and RFI. 4,000 candidates were in the validation set, and another 1,000 independent candidates were used for testing purposes, again split evenly. No data augmentation was necessary. The input data were standardised to be agnostic of DM and observing frequency. The classifier outputs a probability $p_\text{frb} \in [0, 1]$ for each candidate, where $p_\text{frb} = 0.5$ corresponds to a random guess, $p_\text{frb} < 0.5$ indicates RFI, and $p_\text{frb} \geq 0.5$ a pulse. In our tests, \textsc{frbid} achieved an accuracy of $> 99.8$~per cent with a false positive rate of $< 1$~per cent most of the time. The distribution of $p_\text{frb}$ was bimodal with peaks near zero and unity. More details are presented in \citet{2021HosenieThesis}. Finally, we visually inspected the candidates flagged as pulses with $p_\text{frb} \geq 0.5$ by the classifier. We investigated the most promising ones, i.e.\ those identified as astrophysical pulses by both the ML classifier and at least two independent human inspectors, more closely with a custom software tool\footnote{\url{https://bitbucket.org/vmorello/mtcutils/}}. Only those with S/N $\geq 8.0$ as measured in our refined offline analysis that fulfilled strict quality requirements, such having as a well-behaved Gaussian or Lorentzian-like S/N versus trial DM curve and being clearly distinct from RFI, were classified as genuine FRB discoveries.

Unfortunately, the data timestamps reported in this work could, in rare cases, be affected by a known problem in the data processing software. The quoted topocentric arrival times of the FRBs could potentially be earlier than the actual arrival times by exactly one \textsc{psrdada} search block \citep{2021VanStraten}, i.e.\ about 6.115~s here, as a single block could have been skipped. This problem could have also affected earlier work. Aside from this potential offset, the timestamps are known to the precision of the MeerKAT digitizer stage, which is 5~ns \citep{2016Jonas}.

\section{Analysis}
\label{sec:analysis}

\subsection{Scattering fits}

The FRBs presented in this work are of reasonably low S/N, and their data are affected by intra-channel dispersive smearing due to the broad channelisation (1024 channels across 856~MHz of bandwidth, i.e.\ $\sim$0.836~MHz channel bandwidth) for the burst DMs, especially towards the low-frequency band edge. This means that analysing their burst properties is challenging and somewhat limited in scope. For instance, the low S/N prevented us from resolving any scintles in the data, should they exist. Nonetheless, we performed scattering fits to the FRBs using a custom \textsc{python}-based software called \textsc{scatfit}\footnote{\url{https://github.com/fjankowsk/scatfit/}} \citep{2022JankowskiScatfit} in version 0.2.18 that we optimised for low-S/N data. It uses the FRB filterbank data at their native time resolution, robustly estimates model parameters and uncertainties, and the noise present in the profile time series. The observed FRB profile $f$ can be expressed as the convolution product
\begin{equation}
    f(t, \vec{a}) = b + p(t, \vec{a}) \ast s(t, \vec{a}) \ast d(t, \vec{a}) \ast i(t, \vec{a}),
\end{equation}
where $t$ denotes time, $\vec{a}$ is the parameter vector, $b$ is a baseline offset, $p$ is the intrinsic FRB profile, $s$ is the impulse response of the ionised interstellar medium (ISM) and other turbulent ionised media, $d$ is the intra-channel dispersive smearing of the data, $i$ is the instrumental response of the receiver, signal chain and data acquisition system, and $\ast$ denotes linear convolution \citep{2001Loehmer, 2014McKinnon}. We assumed that the FRBs are intrinsically normalised Gaussians
\begin{equation}
    p(t, \vec{a}) = \frac{F}{\sigma \sqrt{2 \pi}} \exp{ \left( -\frac{ \left(t - t_0 \right)^2}{2 \sigma^2} \right)},
\end{equation}
where $F$ is the burst fluence or area under the pulse, $t_0$ is the location parameter, and $\sigma$ is the Gaussian standard deviation that corresponds to the pulse width. For the impulse response of the ionised scattering media $s$ we adopted a single-sided exponential pulse broadening function that characterises isotropic scattering from a thin screen \citep{2001Cordes, 2021Oswald}
\begin{equation}
    s(t, \vec{a}) = \frac{1}{\tau_s} \exp{ \left( - \frac{t}{\tau_s} \right)} H(t),
\end{equation}
where $\tau_s$ is the scatter-broadening time, and $H$ is the Heaviside step function. An approximate value for the intra-channel dispersion smearing due to incoherent dedispersion of a signal of a certain DM is given by
\begin{equation}
    t_\text{dm} = 8.3 \times 10^{-3} \: \text{DM} \: b_\text{c} \: \nu^{-3},
    \label{eq:dmsmearing}
\end{equation}
where $t_\text{dm}$ is in milliseconds, $b_\text{c}$ is the channel bandwidth in MHz, and $\nu$ is the channel centre frequency in GHz. We chose to exclude the DM smearing $d$ and the instrumental impulse response $i$ from the model\footnote{Although \textsc{scatfit} contains more complex profile models that explicitly incorporate the DM smearing and instrumental terms.}, i.e.\ we assumed them to be delta distributions, and instead tested whether the pulse profile fits exceeded the intra-channel DM smearing times. In any case, $i$ was negligible in comparison with $d$ in our data set. The simplified linear convolution can be expressed analytically as an exponentially modified Gaussian
\begin{equation}
    \begin{split}
        f(t, \vec{a}) = b + \frac{F}{2 \tau_s} \: \exp \left( \frac{ \sigma^2 }{ 2 \tau_s^2 } \right) \: & \exp \left( - \frac{t - \mu}{\tau_s} \right) \times\\
        & \text{erfc} \left[ - \frac{1}{\sqrt{2}} \left( \frac{t - \mu}{\sigma} - \frac{\sigma}{\tau_s} \right) \right],
    \end{split}
    \label{eq:frbmodel}
\end{equation}
where $\text{erfc}$ is the complementary error function defined as $\text{erfc} \: (x) = 1 - \text{erf} \: (x)$, and $b \approx 0$ is the baseline offset (slightly adjusted from \citealt{2014McKinnon}). We implemented this analytical approach together with the full numerical convolution model.

We independently fit the scattering model to the cleaned FRB profile data split into several frequency sub-bands and the fully band-integrated data. The data were incoherently dispersed (i.e.\ no coherent dedispersion was applied), as we only have total intensity data for those FRBs and no voltage buffer dumps. We used the \textsc{lmfit} software \citep{2016Newville} to perform initial fits using the Levenberg–Marquardt minimization algorithm \citep{1943Levenberg, 1963Marquardt}. We then used the resulting best-fitting parameters as a starting point for exploring the posteriors using the \textsc{emcee} Markov chain Monte Carlo sampler \citep{2013ForemanMackey}. We constrained the fit parameters to physically reasonable values and ensured that the Markov chains had converged sufficiently using standard techniques. Together with the model parameters, we estimated the standard deviation $\epsilon$ of the noise in the time series data during the sampling process. The software also determined a refined and scattering-corrected DM.

In the following, we report the Gaussian intrinsic pulse widths at 50 and 10 per cent of the maximum, i.e.\ before scattering, smearing, and other instrumental effects, estimated from the $\sigma$ posterior samples as $\text{W}_\text{50i} = 2 \sqrt{2 \ln{(2)}} \: \sigma$ and $\text{W}_\text{10i} = 2 \sqrt{2 \ln{(10)}} \: \sigma$. These are the usual expression for the Gaussian full-width at half (FWHM) and tenth-maximum (FWTM). Additionally, we numerically determine the post-scattering pulse widths $\text{W}_\text{50p}$ and $\text{W}_\text{10p}$ by oversampling (typically by $4 \times$) the resulting FRB profile model for each posterior sample and estimating the 0.16, 0.5, and 0.84 quantiles of the distributions. We do the same for the equivalent pulse width, which is defined as $\text{W}_\text{eq} = \frac{ \sum_i f_i \: \Delta t}{ \max{f} }$, where $f_i$ is the profile amplitude in the $i$-th time sample, $\Delta t$ is the sampling time, and $\max$ denotes the maximum value.

\subsection{Primary and coherent beam models}

\begin{figure}
  \centering
  \includegraphics[width=\columnwidth]{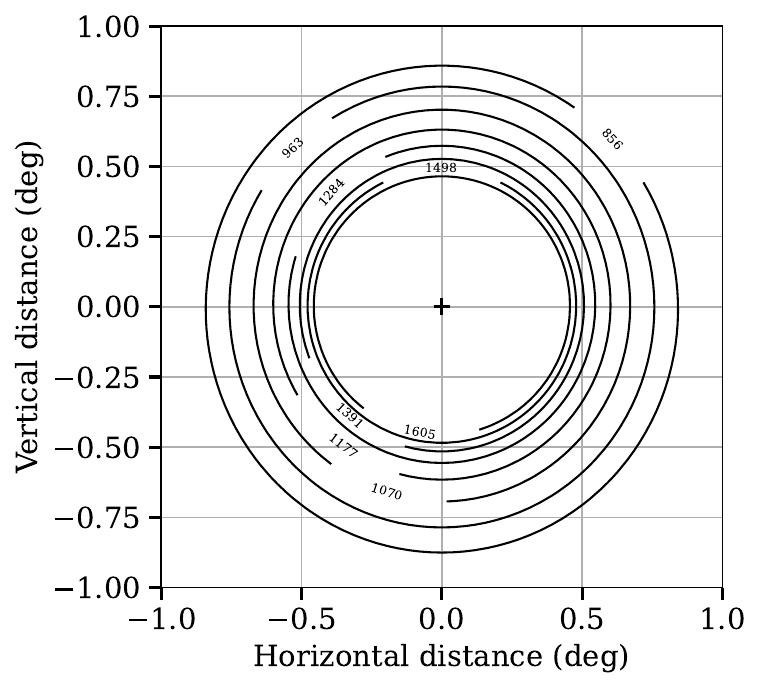}
  \includegraphics[width=\columnwidth]{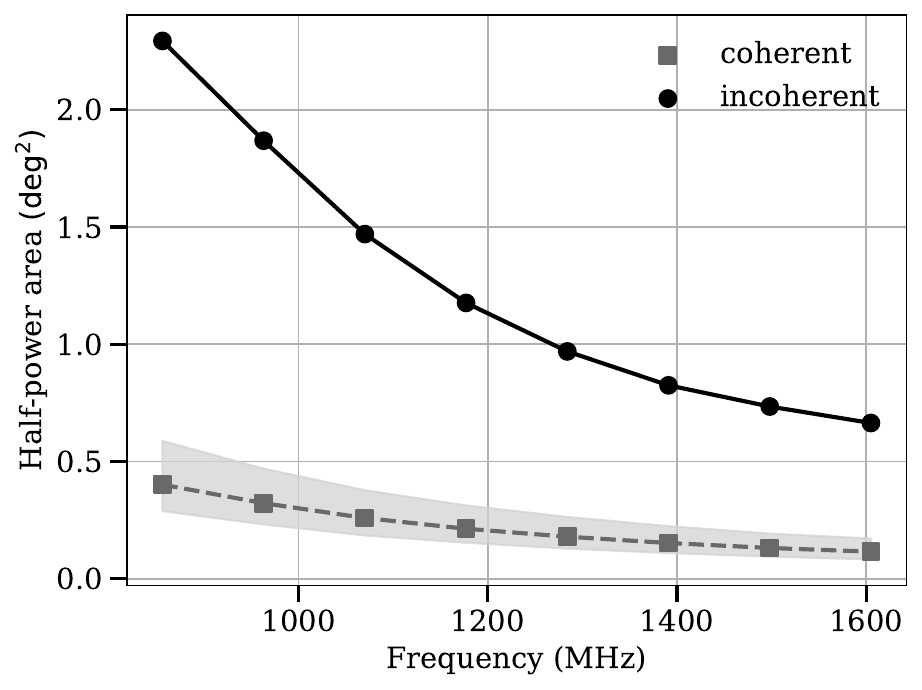}
  \caption{Scaling of the MeerKAT total power (Stokes I) half-power beam area with L-band frequency. Top: Half-power footprints of the MeerKAT primary or incoherent beam in a beam-centred reference frame. The frequencies shown in the plot are in MHz. Bottom: Scaling of the half-power areas with frequency. We show the total half-power area covered with CBs for typically 768 CBs overlapped at the 25 per cent level and the observing configuration used. The areas were computed numerically from the total CB PSF and are corrected for the beam overlap. The shaded area shows the minimum to maximum scatter in the sky area for the FRBs in this paper, while the markers represent the mean values. The total CB area covered varies appreciably with the hour angle and the array configuration.}
 \label{fig:beamarea}
\end{figure}

As we investigate the inferred FRB all-sky rates based on the MeerTRAP survey progress so far in this paper, it is essential to understand the MeerKAT telescope array's beam response accurately. For the IB, we based our analysis on astro-holographic measurements of the MeerKAT Stokes I primary beam response at L-band \citep{2021Asad, 2022DeVilliers}, which are consistent with a cosine-tapered field illumination pattern at small radial distances \citep{2020Mauch}. The cosine-tapered primary beam parameterisation is available in the \textsc{katbeam} \textsc{python} package\footnote{\url{https://github.com/ska-sa/katbeam/}}. For the CBs, we numerically simulated the individual CB PSF and the beam tiling pattern on the sky using the \textsc{mosaic} beam synthesis code \citep{2021Chen} for typical MeerTRAP observing configurations. We then computed the total aggregate CB response by reprojecting the individual CB PSF to each tiling location in the grid. The total CB response is the superposition of the individual PSFs and the survey coverage is given by the maximum value over all contributing individual PSFs for each pixel in the total CB PSF array. We calculated the total half-power area from this by summing the sky area of at least 0.5 in the total CB response. Hence, the resulting estimate is corrected for (i.e.\ excludes) the beam overlap by design. The beam overlap at half-power, meaning the difference between our total area values and the simple CB area sum $N_\text{beam} \times A_{0.5}(\text{1 CB})$, where $N_\text{beam}$ is the number of CBs in the grid and $A_{0.5}(\text{1 CB})$ is the half-power area of an individual CB, amounted to at most 1 per cent for the FRBs in this paper. This is unsurprising, as the CBs were spaced relatively far apart (0.25 level) so that their half-power areas do not significantly overlap except in pathological cases, e.g.\ at extreme hour angles or low frequencies. In Fig.~\ref{fig:beamarea} we show the scaling of the IB half-power area with frequency, and we compare the half-power beam areas of the IB and the total area tesselated with CBs.

\subsection{FRB tied-array beam localisation}

Each FRB presented in this work was detected in only a single CB in a tiling wherein the CBs overlapped at 25 per cent of their maximum sensitivity. Hence, it is difficult to constrain their positions much more precisely than an ellipse fit to the 25 per cent level of the detection CB point spread function (PSF). However, by considering the non-detection of each pulse in neighbouring beams, one can add additional constraints. To do this, we modelled the MeerKAT CB PSFs using the \textsc{mosaic} software and arranged the PSFs to correspond to the centre coordinates of all the CBs formed during the observation. Considering that the detection threshold for the MeerTRAP single-pulse search pipeline was S/N 8.0, we then used the beam models to determine how close the FRB could have been to a neighbouring CB without having been detected with a $\text{S/N} > 8.0$. All viable positions for the FRB must comply with the relation
\begin{equation}
    \frac{S_\text{det}}{S_i} \geq \frac{\text{S/N}_\text{det}}{\text{S/N}_\text{thresh}},
    \label{eq:localisation}
\end{equation}
where $S_\text{det}$ is the PSF of the detection CB, $\text{S/N}_\text{det}$ is the S/N of the detection, $S_i$ is the PSF of each other CB, and $\text{S/N}_\text{thresh}$ is the detection threshold S/N. Coordinates fulfilling Eq.~\ref{eq:localisation} were assigned a value of unity, while all others were assigned a value of zero to produce a ``localisation mask''. The localisation probability density function (PDF) was then taken as the localisation mask, normalised such that the sum over all sky area equalled unity, i.e.\ a uniform PDF within the localisation region. This is a conservative approach and is agnostic of FRB population or FRB detection rate parameters. Allowing for an uncertainty of unity in $\text{S/N}_\text{det}$ negligibly changes the localisation regions. We estimated a relative change in localisation area of $\leq 1$~per cent for the FRBs in this paper. For more details about our tied-array beam localisation method named ``TABLo'', see \citet{2023Bezuidenhout}. We implemented the technique in a \textsc{python}-based software called \textsc{seekat}\footnote{\url{https://github.com/BezuidenhoutMC/SeeKAT/}}.

\subsection{\textsc{path} FRB host galaxy association}

Based on our best localisations of the FRBs, we used the Probabilistic Association of Transients to their Hosts (\textsc{path}; \citealt{2021Aggarwal}) software to assign each galaxy detected within the localisation region a probability of being the FRB's host. To do that, we first generated high-resolution \textsc{healpix} localisation maps \citep{2005Gorski} that \textsc{path} can read directly. That was necessary because the localisation regions are complex in shape. The probability density is uniform within those regions and vanishes outside. We then retrieved our chosen optical catalogue data for that field, as described separately for each FRB below, and selected the sources' centroid positions, apparent magnitudes, and half-light radii, usually for the \textit{i}-band data. For reference, the \textit{i}-band filter of the Pan-STARRS1 (PS1) photometric system has an approximate square-bandpass response between 690 and 819~nm \citep{2012Tonry, 2016Chambers}. We reduced our selection to the sources that were clearly extended beyond the PSF of the image, i.e.\ the galaxies, and that simultaneously fulfilled strict data quality requirements. Throughout the analysis, we assumed the default \textsc{path} priors with one modification, i.e.\ that the candidate probability scales inversely proportional to the sky density of galaxies with that apparent magnitude (inverse prior; brighter candidates have higher prior probability), a zero probability that the true host is unseen in the image, and an exponential prior with a scale of 0.5 on the FRB's offset from the candidate galaxy's optical centroid and truncated at six times its angular size (inner galaxy regions have higher probability). The adjustment to the scale was motivated by the observed offset distribution of well-localized FRBs (Shannon et al.\ in prep.). For the FRB fields with shallow optical coverage, we cannot exclude that faint host galaxy candidates were undetected within the localisation regions. We estimated the probability of an undetected host by artificially adding ten mock galaxies to the \textsc{path} analysis. We randomly distributed them within the localisation regions and set their half-light radii to $2 \arcsec$ and their apparent magnitudes to one mag above the faintest galaxy detected in the region. The numbers are conservative and were chosen to approximately match the number of galaxies detected in the FRB field with deep optical imaging data -- FRB~20201211A. In comparison with the candidates in that field, the mock galaxies are brighter than all candidates for FRB~20210202D and brighter than about half for FRB~20210408H. They are less extended than all of them. The total prior and posterior probabilities for an unseen host, which we denote as $p(M)$ and $p(M|x)$, are the sums over the mock galaxies' prior and posterior probabilities.

\subsection{Expected host galaxy redshifts}

When investigating FRB host galaxy candidates, it is useful to know the expected redshift range. To do that, we computed the redshift ranges based on the FRBs' cosmic DMs. The observed FRB DM, which we denote simply as DM, is the sum of the DM contributions from the Galactic ISM, Milky Way halo, IGM, intervening galaxies and halos, and the FRB's host galaxy and halo as
\begin{equation}
    \begin{split}
    \text{DM} = \text{DM}_\text{mw} + \text{DM}_\text{halo} + \text{DM}_\text{igm} (z) & + \sum_j \frac{ \text{DM}_{\text{igh},j} }{ 1 + z_{\text{igh},j} }\\
    & + \frac{ \text{DM}_\text{host} }{ 1 + z },
    \end{split}
    \label{eq:totaldm}
\end{equation}
where $z$ is the host galaxy redshift and $z_{\text{igh}}$ is the redshift of an intervening galaxy or its halo. The cosmic DM is the sum
\begin{equation}
    \text{DM}_\text{cosmic} (z) = \text{DM}_\text{igm} (z) + \sum_j \frac{ \text{DM}_{\text{igh},j} }{ 1 + z_{\text{igh},j} }.
    \label{eq:cosmicdm}
\end{equation}
As several of those DM contributions are poorly known, we define an FRB's extragalactic DM as
\begin{equation}
    \text{DM}_\text{eg} = \text{DM} - \text{DM}_\text{mw} - \text{DM}_\text{halo}.
\end{equation}
That is because we have more established models for the Galactic contributions. In particular, for $\text{DM}_\text{mw}$, we used the mean value of the ISM contributions computed using the \textsc{ne2001} \citep{2002Cordes} and \textsc{ymw16} \citep{2017Yao} Galactic free-electron models. For the Milky Way halo contribution, we assumed the \citet{2020Yamasaki} model. We neglected the $\text{DM}_\text{igh}$ term in Eq.~\ref{eq:totaldm}-\ref{eq:cosmicdm}, as we do not know exactly what galaxies (and haloes) an FRBs traversed. As the host galaxy DM contribution is a matter of active research, we assumed a uniform distribution between 30 and $300~\text{pc} \: \text{cm}^{-3}$ for it in the observer's reference frame, informed by the spread in currently-known FRB hosts. This choice covers all but one published localised burst, FRB~20190520B, which has an unexpectedly large host DM contribution of $\sim$900~$\text{pc} \: \text{cm}^{-3}$ \citep{2022Niu}. Finally, we performed the cosmological integration from cosmic DM to host galaxy redshift using the \textsc{fruitbat} software \citep{2019Batten} while assuming the ``Planck 2018'' cosmology \citep{2020Planck} and the \citet{2018Zhang} cosmic DM to redshift relation.

The above treatment assumed that the primary redshift uncertainty lay in the host galaxy DM contributions and neglected any spread around the cosmic DM -- redshift relation. Hence, we additionally used an alternative approach\footnote{\url{https://github.com/FRBs/FRB/}} to account for it that evaluates the \citet{2020Macquart} relation and its spread and estimates the probability of a host galaxy redshift given an observed DM, i.e.\ $p(z|\text{DM})$. The software uses the \textsc{ne2001} model for the ISM contribution, and we fixed the combined Milky Way halo and host galaxy DM contribution to $100~\text{pc} \: \text{cm}^{-3}$.

\section{Results}
\label{sec:results}

\subsection{An FRB sample discovered with MeerTRAP}
\label{sec:frbsample}

\begin{figure}
  \centering
  \includegraphics[width=\columnwidth]{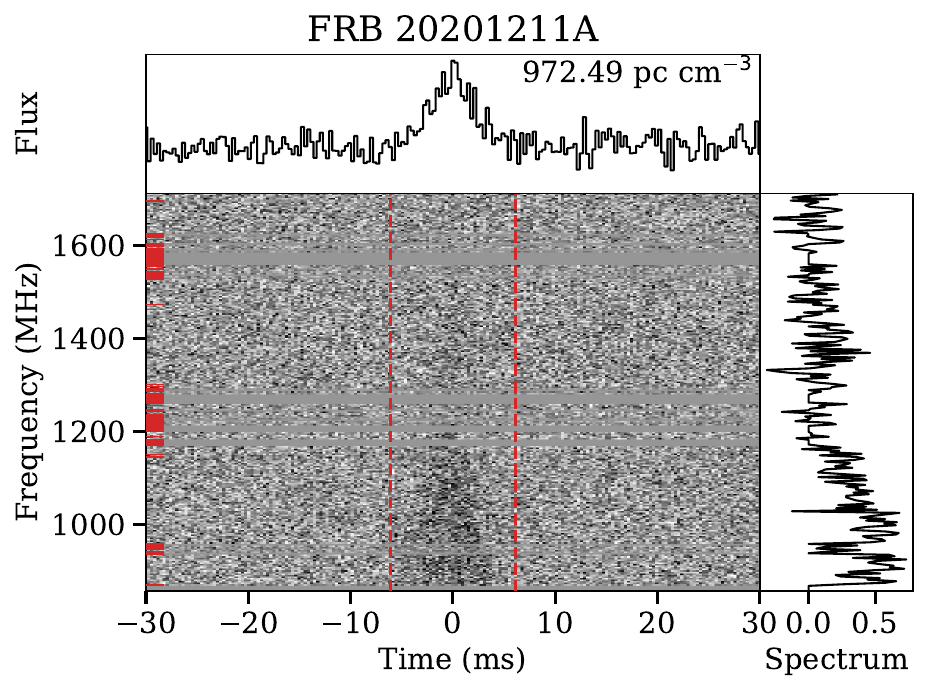}
  \includegraphics[width=\columnwidth]{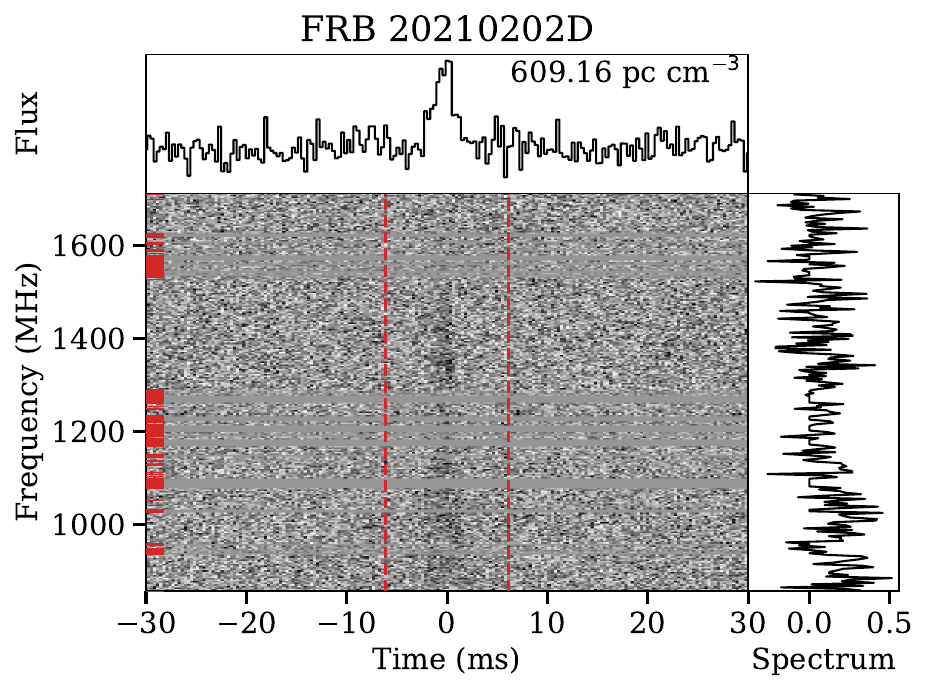}
  \includegraphics[width=\columnwidth]{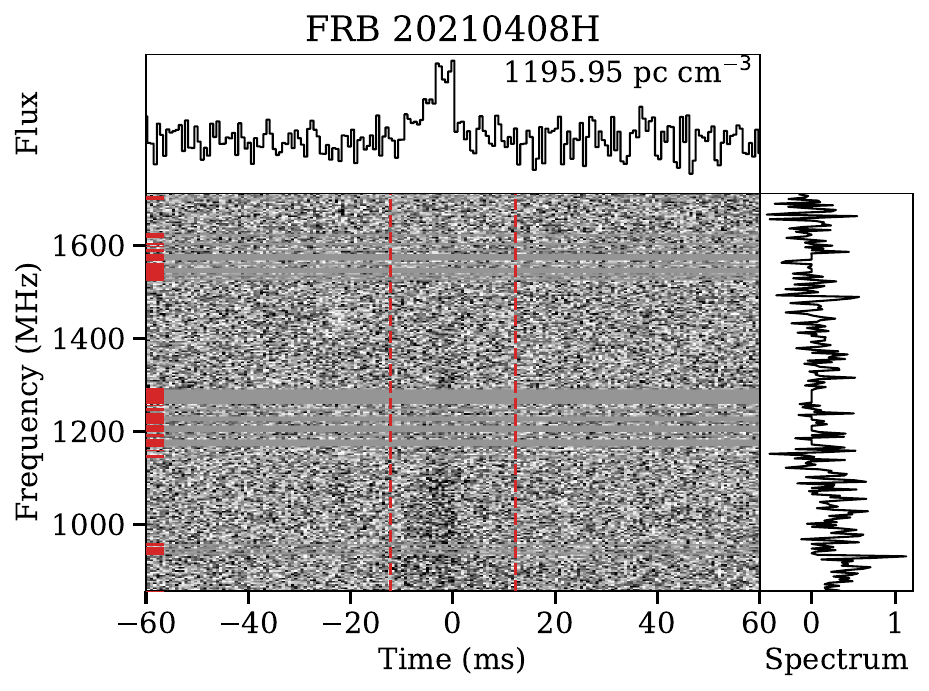}
  \caption{Dedispersed dynamic spectra, pulse profiles and uncalibrated total intensity spectra of the FRBs presented in this work. The data were dedispersed at the best-determined S/N-optimising DMs shown in the top right corners. We summed every four frequency channels for clarity. For FRB~20210408H, we additionally summed every two time samples; the others are displayed at the native time resolution of the data. The horizontal red lines indicate the native frequency channels that were masked, and the vertical, dashed red lines highlight the on-burst regions from which the spectra were computed.}
 \label{fig:burstplots}
\end{figure}

\begin{table*}
\caption{Properties of the FRBs presented in this work. We list their measured parameters, i.e.\ their topocentric arrival times, barycentric arrival times referenced to an infinite frequency in the Barycentric Dynamical Time (TDB) time scale, detection beams, best-determined ICRS positions, Galactic coordinates, S/N-optimising DMs, S/N, Gaussian intrinsic and post-scattering pulse widths at 50 and 10 per cent maximum, scattering times $\tau_s$, and boxcar equivalent widths. Additionally, we present their instrumental properties like DM smearing times, numbers of frequency channels, effective bandwidths after RFI excision, MeerKAT antennas in the coherent and incoherent sums, sky areas covered by the detection CBs and localisation regions, angular separations from the boresight, the accumulated observing time on the FRB fields up to the end of 2021, and the MeerKAT Large Survey Projects that MeerTRAP were commensal with at the times of discovery. The inferred parameters are their peak flux densities $S_\text{peak}$, fluences $F$, Galactic and Milky Way halo DM contributions, extragalactic DMs, and the expected host galaxy redshift ranges.}
\label{tab:burstproperties}
\begin{tabular}{lcccc}
\hline
FRB	                    &                                   & 20201211A & 20210202D    & 20210408H\\
Parameter               & Unit\\
\hline
\multicolumn{5}{c}{Measured}\\
MJD$_\text{topo}^\text{a}$  &                               & 59194.894135696   & 59247.526682167 & 59312.889025614\\
UTC$_\text{topo}^\text{a}$  &                               & 2020-12-11 21:27:33.324    & 2021-02-02 12:38:25.339    & 2021-04-08 21:20:11.813\\
MJD$_\text{bary}^\text{b}$  &                               & 59194.898442783   & 59247.523099300     & 59312.895193659\\
Beam                    &                                   & 305C  & 337C  & 357C\\
RA$^\text{c}$           & (hms)                             & 04:29:45.51 $\pm 2.6s$    & 19:46:48.74 $\pm 8.7s$ & 13:37:18.25 $\pm 3.3s$\\
Dec$^\text{c}$          & (dms)                             & $-$27:30:28.3 $\pm 41s$   & $-$54:13:38.8 $\pm 58s$   & $-$28:17:02.9 $\pm 50s$\\
l                       & (deg)                             & 226.666713    & 343.699648    & 315.051287\\
b                       & (deg)                             & $-$41.920362  & $-$29.633060  & 33.505919\\
DM                      & ($\text{pc} \: \text{cm}^{-3}$)   & $972.49 \pm 0.95$ & $609.16 \pm 0.57$ & $1195.95 \pm 1.5$\\
$\text{S/N}^\text{d}$   &                                   & 25.4  & 18.6  & 14.7\\
W$_\text{50i}^\text{e}$ & (ms)                              & $4.6 \pm 0.5$ & $2.2 \pm 0.3$ & $6.1 \pm 0.9$\\
W$_\text{10i}^\text{e}$ & (ms)                              & $8.3 \pm 0.9$ & $4.0 \pm 0.5$ & $11.1 \pm 1.6$\\
W$_\text{50p}^\text{e}$ & (ms)                              & $4.9 \pm 0.3$ & $2.3 \pm 0.2$ & $6.4 \pm 0.9$\\
W$_\text{10p}^\text{e}$ & (ms)                              & $9.1 \pm 0.6$ & $4.4 \pm 0.4$ & $11.6 \pm 1.7$\\
$\tau_s^\text{e}$       & (ms)                              & $1.0 \pm 0.6$ & $0.5 \pm 0.3$ & $1.0 \pm 0.8$\\
W$_\text{eq}^\text{e}$  & (ms)                              & $5.3 \pm 0.4$ & $2.6 \pm 0.3$ & $6.8 \pm 0.9$\\
\hline
\multicolumn{5}{c}{Instrumental}\\
$t_\text{dm}^\text{f}$  & (ms)                              & 10.7  & 6.7   & 13.2\\
$N_\text{chan}$         &                                   & 1024  & 1024  & 1024\\
$b_\text{eff}$          & (MHz)                             & 679.7 & 644.6 & 668.5\\
$N_\text{ant,ib}$       &                                   & 62    & 55    & 64\\
$N_\text{ant,cb}$       &                                   & 44    & 32    & 44\\
$A_\text{cb}$           & (arcmin$^2$)                      & 0.7   & 1.4   & 0.7\\
$A_\text{loc}$          & (arcmin$^2$)                      & 1.2   & 2.9   & 1.7\\
$\delta_\text{bore}^\text{g}$   & (deg)                     & 0.101 & 0.169 & 0.239\\
$a_\text{IB}$           &                                   & 0.980 & 0.942 & 0.886\\
$t_\text{obs}$          & (h)                               & 26.8  & 4.9   & 22.0\\
LSP                     &                                   & MHONGOOSE & MeerTime (MSP)    & MHONGOOSE\\
\hline
\multicolumn{5}{c}{Inferred}\\
$S_\text{peak}$     & (mJy)                             & $> 110$   & $> 169$   & $> 63$\\
$F$                 & (Jy ms)                           & $> 0.6$   & $> 0.4$   & $> 0.4$\\
DM$_{\text{mw,n}}$  & ($\text{pc} \: \text{cm}^{-3}$)   & 39  & 69  & 57\\
DM$_{\text{mw,y}}$  & ($\text{pc} \: \text{cm}^{-3}$)   & 37  & 51  & 43\\
DM$_{\text{halo}}$  & ($\text{pc} \: \text{cm}^{-3}$)   & 32  & 64  & 48\\
DM$_\text{eg}$      & ($\text{pc} \: \text{cm}^{-3}$)   & 903 & 486 & 1098\\
$z$                 &                                   & [0.675, 0.968]    & [0.219, 0.516]    & [0.886, 1.184]\\
\hline
\multicolumn{5}{l}{$^\text{a}$ At the highest frequency channel, 1711.58203125~MHz. The topocentric arrival times could potentially be}\\
\multicolumn{5}{l}{earlier than the actual arrival times by exactly one \textsc{psrdada} search block, i.e.\ about 6.115~s.}\\
\multicolumn{5}{l}{$^\text{b}$ Barycentric burst arrival time referenced to infinite frequency.}\\
\multicolumn{5}{l}{$^\text{c}$ We quote half the maximum projected extents of the localisation regions as uncertainties, i.e. areas}\\
\multicolumn{5}{l}{larger than their circumellipses. Region files and \textsc{healpix} localisation maps are available online.}\\
\multicolumn{5}{l}{$^\text{d}$ After RFI excision, DM, and pulse width refinement, not discovery S/N.}\\
\multicolumn{5}{l}{$^\text{e}$ At the centre of the band, 1284~MHz.}\\
\multicolumn{5}{l}{$^\text{f}$ Intra-channel dispersive smearing in the lowest frequency channel.}\\
\multicolumn{5}{l}{$^\text{g}$ Angular separation of the centre of the detection CB from the boresight pointing.}\\
\end{tabular}
\end{table*}

\begin{figure*}
  \centering
  \includegraphics[width=0.49\textwidth]{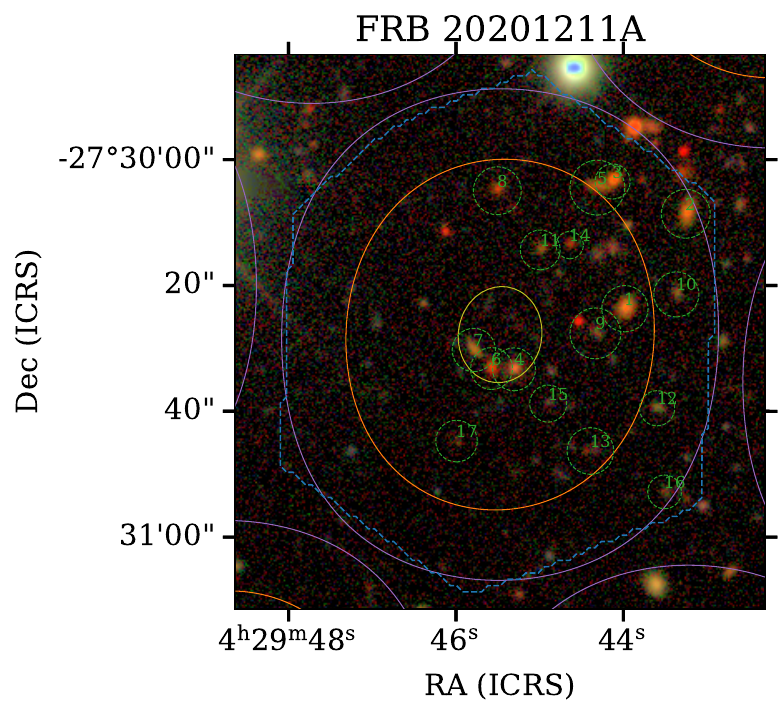}
  \includegraphics[width=0.49\textwidth]{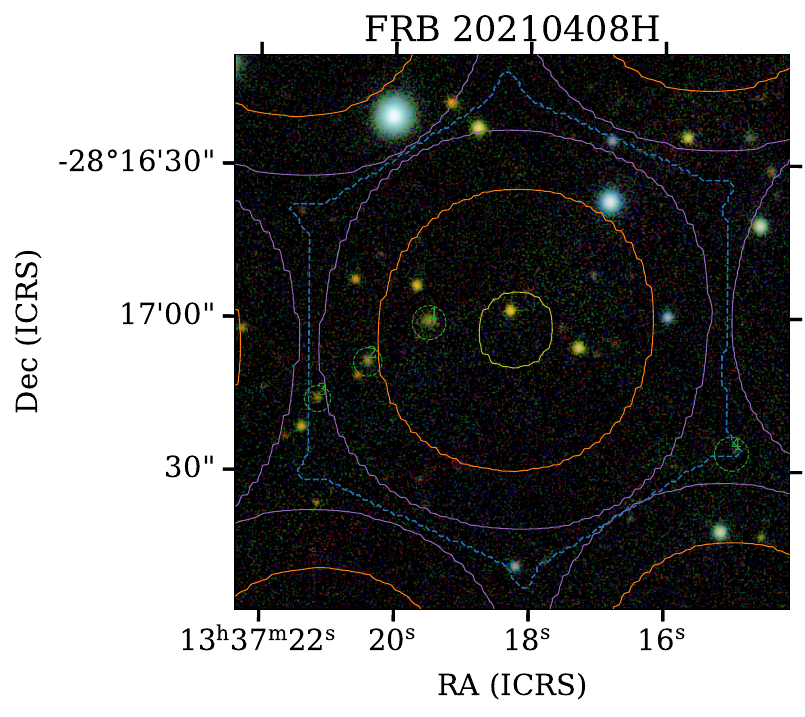}
  \includegraphics[width=0.49\textwidth]{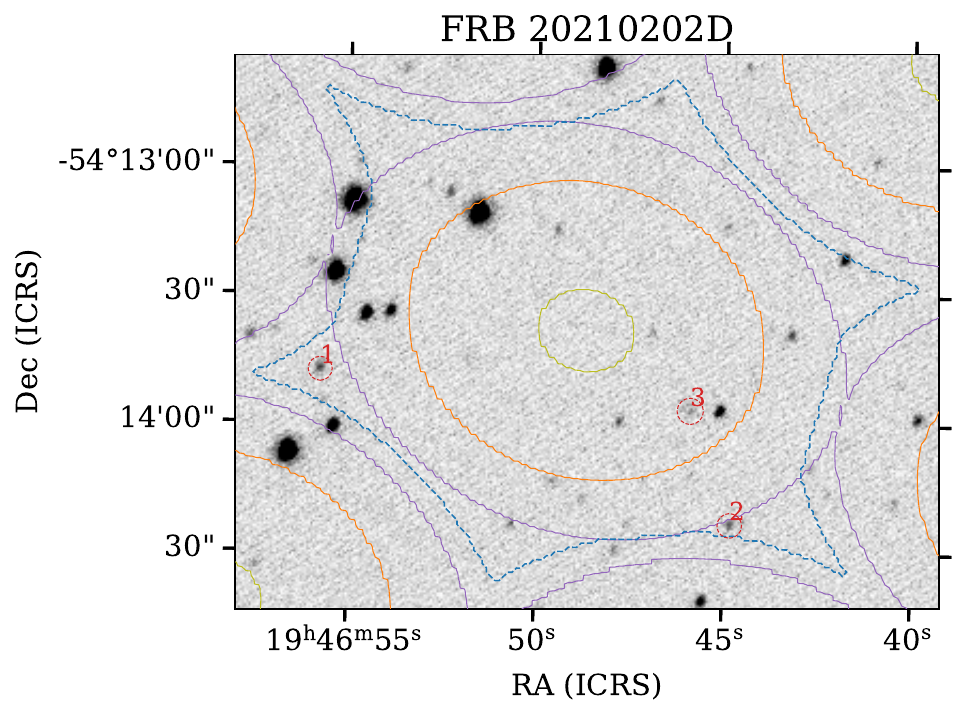}
  \caption{The best localisations of the FRBs presented in this work in the context of optical observations of the discovery fields. We show from top left to bottom: FRBs~20201211A, 20210408H, and 20210202D. The background raster images show DES DR2 \textit{irg}-band (FRB~20201211A), Pan-STARRS1 DR1 \textit{zig}-band (FRB~20210408H), and SkyMapper DR2 \textit{i}-band (FRB~20210202D) optical imaging data. We marked the 25, 50, and 95 per cent level contours of the total coherent beam PSF with solid purple, orange, and olive lines. The localisation regions are shown with dashed blue lines and the host galaxy candidates ranked by their posterior probabilities with green or red dashed lines.}
 \label{fig:localisations}
\end{figure*}

\begin{table}
\caption{Results of the host galaxy association (\textsc{path}) analysis for the FRBs presented in this work. We list the ranked galaxy numbers, the posterior $p(O|x)$ and prior $p(O)$ probabilities, the centroid positions, the apparent \textit{i}-band magnitudes $m_i$, and the half-light radii $\phi$. The magnitudes in the FRB~20201211A and FRB~20210408H fields were de-reddened. We assumed no unseen galaxies for the deep optical data of the FRB 20201211A field but added ten unseen mock galaxies in the two other fields with shallower optical coverage. $p(M)$ and $p(M|x)$ are the mock galaxies' total prior and posterior probabilities.}
\label{tab:pathanalysis}
\begin{tabular}{lcccccc}
\hline
\#    & $p(O|x)$    & $p(O)$    & RA    & Dec   & $m_i$ & $\phi$\\
      &             &           & (deg) & (deg) & (mag) & (\arcsec)\\
\hline
\multicolumn{7}{c}{FRB~20201211A}\\
\multicolumn{7}{c}{$p(M)$ = 0, $p(M|x)$ = 0}\\
1     & 0.174   & 0.167   & 67.43326      & $-$27.50658     & 20.74   & 3.68\\
2     & 0.154   & 0.164   & 67.43023      & $-$27.50239     & 20.75   & 3.89\\
3     & 0.132   & 0.142   & 67.43387      & $-$27.50097     & 20.91   & 2.79\\
4     & 0.101   & 0.101   & 67.43878      & $-$27.50926     & 21.26   & 3.41\\
5     & 0.093   & 0.082   & 67.43463      & $-$27.50124     & 21.48   & 4.36\\
6     & 0.061   & 0.061   & 67.43989      & $-$27.50922     & 21.79   & 3.33\\
7     & 0.058   & 0.057   & 67.44078      & $-$27.50840     & 21.87   & 3.44\\
8     & 0.051   & 0.048   & 67.43961      & $-$27.50137     & 22.07   & 3.81\\
9     & 0.032   & 0.030   & 67.43473      & $-$27.50767     & 22.61   & 4.04\\
10    & 0.029   & 0.029   & 67.43070      & $-$27.50595     & 22.63   & 3.59\\
11    & 0.024   & 0.025   & 67.43747      & $-$27.50399     & 22.80   & 3.17\\
12    & 0.021   & 0.023   & 67.43166      & $-$27.51098     & 22.92   & 2.79\\
13    & 0.019   & 0.018   & 67.43498      & $-$27.51286     & 23.16   & 3.71\\
14    & 0.015   & 0.016   & 67.43600      & $-$27.50377     & 23.32   & 2.20\\
15    & 0.012   & 0.013   & 67.43708      & $-$27.51077     & 23.58   & 2.94\\
16    & 0.012   & 0.013   & 67.43128      & $-$27.51467     & 23.56   & 2.68\\
17    & 0.011   & 0.011   & 67.44164      & $-$27.51243     & 23.82   & 3.31\\
\hline
\multicolumn{7}{c}{FRB~20210202D}\\
\multicolumn{7}{c}{$p(M)$ = 0.4, $p(M|x)$ = 0.45}\\
1     & 0.246   & 0.238   & 296.73226     & $-$54.22996     & 18.71   & 2.80\\
2     & 0.173   & 0.188   & 296.68669     & $-$54.23982     & 18.93   & 2.87\\
3     & 0.135   & 0.125   & 296.69120     & $-$54.23245     & 19.31   & 3.07\\
\hline
\multicolumn{7}{c}{FRB~20210408H}\\
\multicolumn{7}{c}{$p(M)$ = 0.3, $p(M|x)$ = 0.34}\\
1     & 0.351   & 0.308   & 204.33124     & $-$28.28362     & 20.12   & 3.32\\
2     & 0.172   & 0.160   & 204.33501     & $-$28.28578     & 20.77   & 2.79\\
3     & 0.094   & 0.110   & 204.33810     & $-$28.28780     & 21.16   & 2.59\\
4     & 0.040   & 0.084   & 204.31249     & $-$28.29069     & 21.45   & 3.34\\
\hline
\end{tabular}
\end{table}

We present three FRBs discovered with MeerTRAP at L-band, all of which are localised to a single tied-array beam, i.e.\ to about $1~\text{arcmin}^2$ or better. We list their burst properties in Tab.~\ref{tab:burstproperties} and show their dedispersed dynamic spectra, pulse profiles and uncalibrated total intensity spectra in Fig.~\ref{fig:burstplots}. As DM uncertainties, we quote the half-range for which the S/N versus trial DM curve dropped by unity combined in quadrature with the (somewhat smaller) error from the DM refinement in the scattering fit. For the positional uncertainties, we state half the maximum projected extents of the localisation regions, which are larger and therefore more conservative than their circumellipses. The quoted localisation areas $A_\text{loc}$ give an accurate representation. We used the JPL DE440 Solar System ephemeris \citep{2021Park} to convert the topocentric to barycentric arrival times at infinite frequency. In Fig.~\ref{fig:localisations} we show their best localisations on the sky in reference to optical imaging data of their discovery fields. The blue contours delineate the furthest viable coordinates from the detection beam centres that comply with Eq.~\ref{eq:localisation}. Using our ``TABLo'' technique, we localised the FRBs to asymmetric regions slightly larger than the half-power points of the detection beams, as shown with orange ellipsoids, which typically are smaller than $1~\text{arcmin}^2$. The localisations are available as region files and high-resolution \textsc{healpix} sky maps from our Zenodo repository. The false colour raster images were produced using the \citet{2004Lupton} algorithm. Tab.~\ref{tab:pathanalysis} reports the \textsc{path} probabilities of the galaxies marked in Fig.~\ref{fig:localisations}. We discuss each FRB in the following.

\subsubsection{FRB~20201211A}

\begin{figure*}
  \centering
  \includegraphics[width=0.49\textwidth]{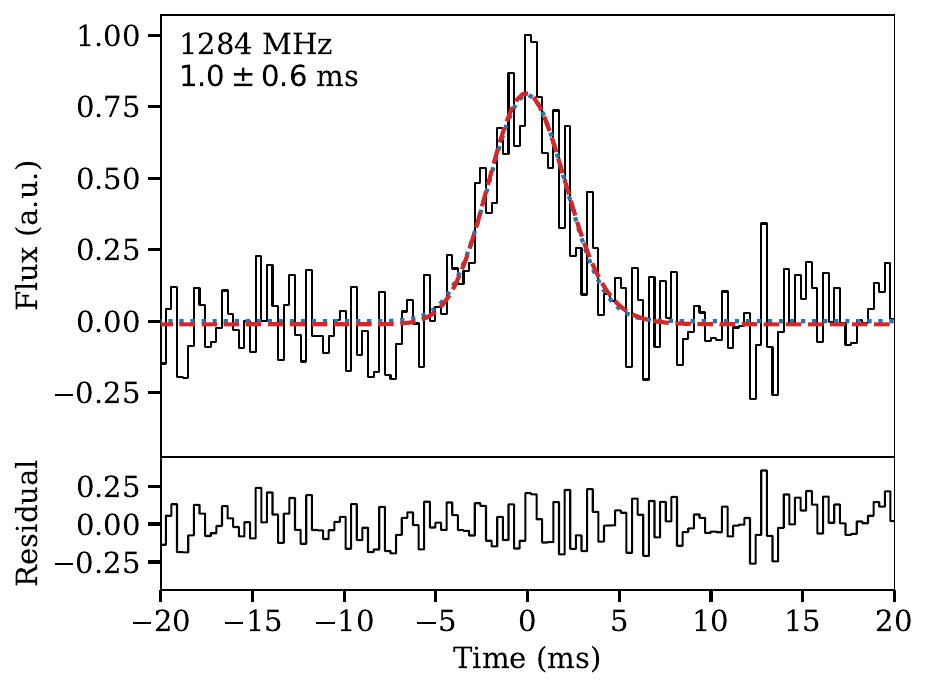}
  \includegraphics[width=0.49\textwidth]{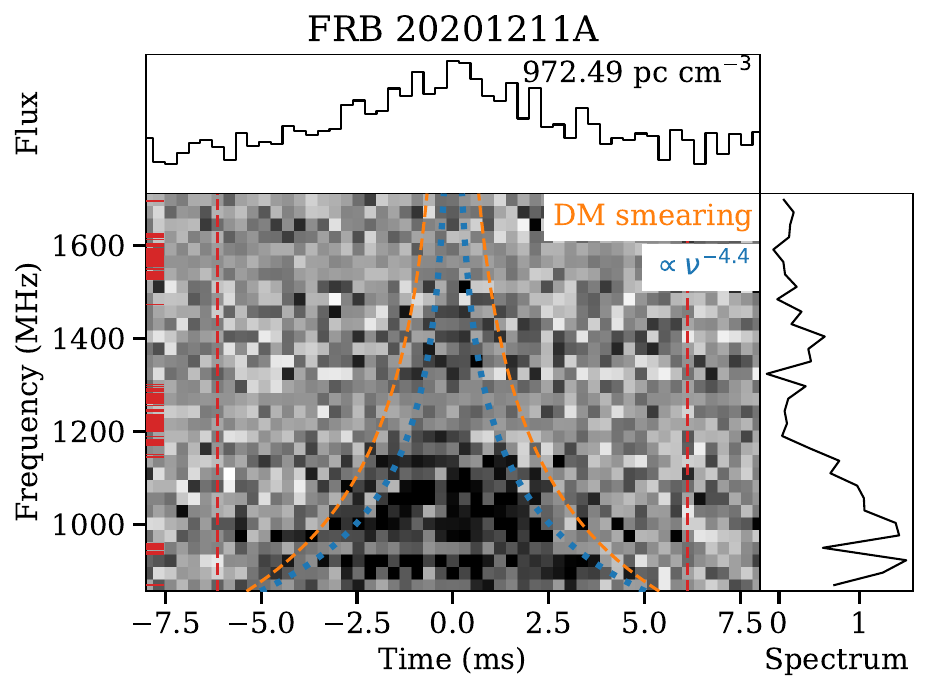}
  \includegraphics[width=0.49\textwidth]{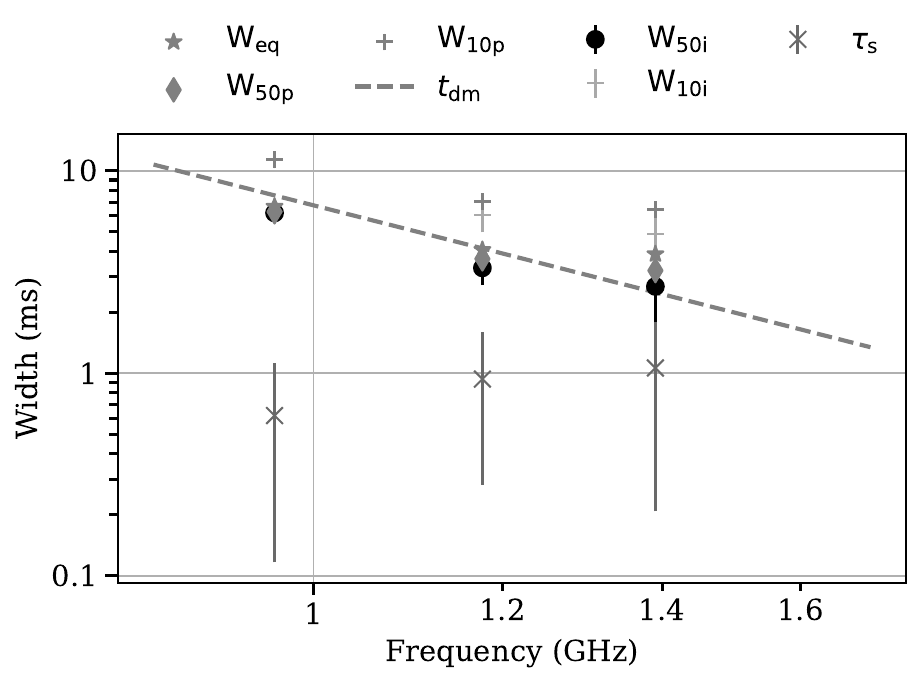}
  \includegraphics[width=0.49\textwidth]{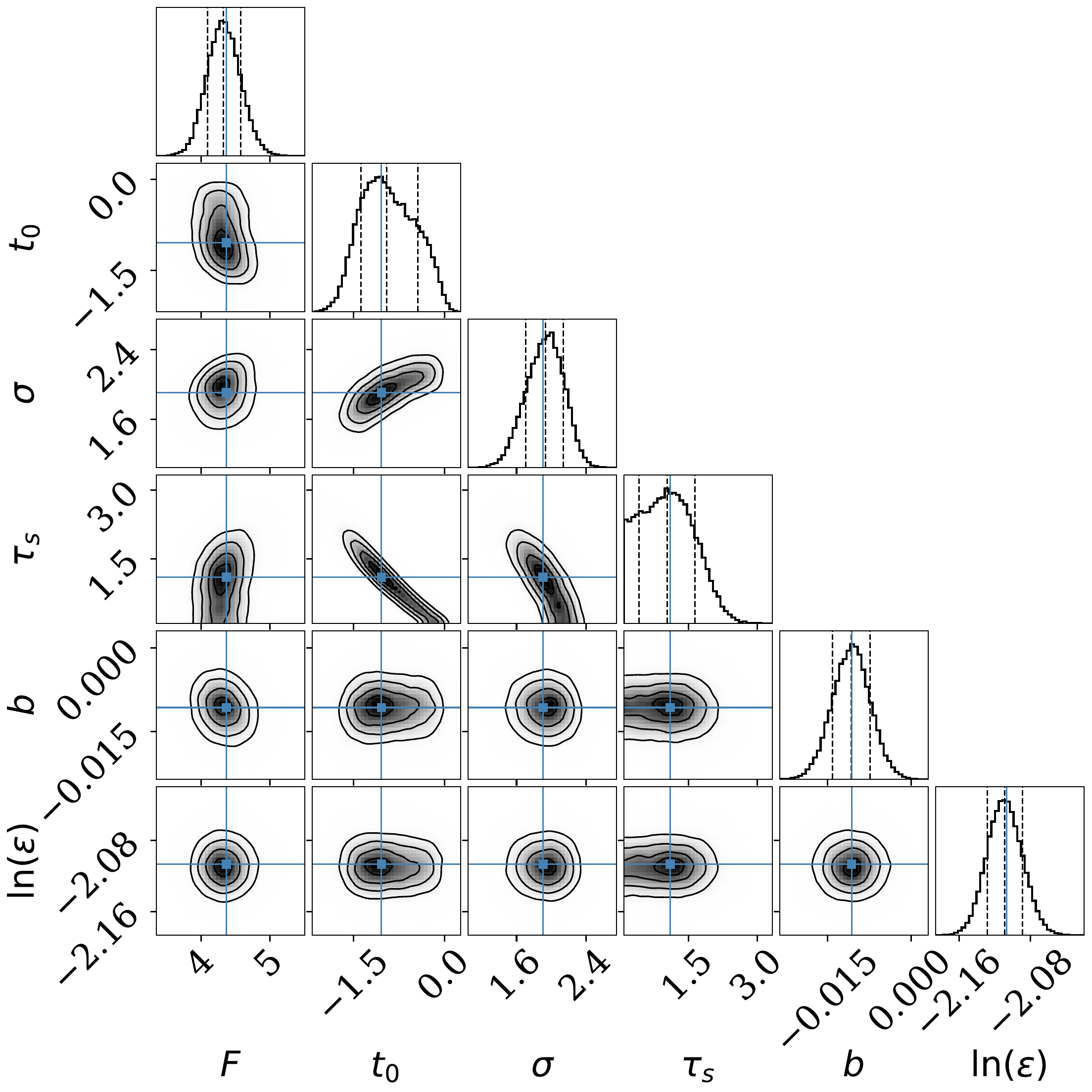}
  \caption{Scattering analysis of FRB~20201211A that is representative for all the FRBs presented here. Top left: Dedispersed band-integrated profile at the native time resolution of the data with our best scattering fit overlaid (red dashed line) in comparison with the best-fitting unscattered Gaussian model (blue dotted line). The fits are almost identical. Top right: Dedispersed dynamic spectrum showing the scaling of the pulse width with frequency. We display the expected instrumental pulse broadening $t_\text{dm}$ due to intra-channel DM smearing (Eq.~\ref{eq:dmsmearing}), and scattering in the ISM and other turbulent ionised media assuming Kolmogorov turbulence $\propto \nu^{-4.4}$, where $\nu$ is the observing frequency. Bottom left: Scaling of the best-fitting Gaussian intrinsic pulse widths $\text{W}_\text{50i}$ and $\text{W}_\text{10i}$, observed post-scattering pulse widths $\text{W}_\text{50p}$ and $\text{W}_\text{10p}$, equivalent widths $\text{W}_\text{eq}$, and scatter-broadening times $\tau_s$ with frequency, in comparison with the expected DM smearing. Bottom right: A corner plot corresponding to the scattering fit in the top left panel. The blue squares mark the maximum likelihood values and the dashed lines the medians and the 68~per cent credibility ranges. The marginalised $\tau_s$ posterior peaks at small but non-zero values. The FRB is unresolved in our data, and its pulse broadening is largely consistent with instrumental DM smearing.}
 \label{fig:scattering}
\end{figure*}

We discovered FRB~20201211A in data taken on 2020-12-11 UTC of the NGC1592 field, on which MeerTRAP was commensal with the ``MeerKAT HI Observations of Nearby Galactic Objects -- Observing Southern Emitters'' (MHONGOOSE) LSP \citep{2016DeBlok} that studies the neutral hydrogen content of nearby galaxies. The FRB has a S/N of 25.4, a S/N-optimising DM of $972.49~\text{pc} \: \text{cm}^{-3}$, an extragalactic DM of $903~\text{pc} \: \text{cm}^{-3}$, a $\text{W}_\text{50p}$ pulse width of 4.9~ms, an inferred peak flux density $> 110~\text{mJy}$, and a fluence $> 0.6~\text{Jy}~\text{ms}$. The expected host galaxy redshift range is [0.675, 0.968] when assuming a uniform distribution of host galaxy DM. For a combined and fixed Milky Way halo and host galaxy contribution of $100~\text{pc} \: \text{cm}^{-3}$ and taking into account the uncertainty in the \citet{2020Macquart} DM -- redshift relation, the expected host galaxy redshift range is [0.543, 1.307] at the 95~per~cent confidence level. There are extant optical imaging data for this field from the Dark Energy Survey Data Release 2 \citep[DES DR2;][]{2021Abbot} with approximate limiting magnitudes of 24.7 (\textit{g}) and 23.8 (\textit{i}-band). We used the source data provided by the DES DR2 team for that field in the \textit{i}-band filter, i.e.\ the source centroid locations, de-reddened apparent magnitudes and half-light radii as input for the \textsc{path} software. We only selected those sources that are galaxies with high confidence (\texttt{extended class} flags) and enforced strict quality requirements on the input candidates (\texttt{imaflag iso}, \texttt{flags}, \texttt{niter model}, and \texttt{nepochs} flags). Out of those, we selected the host galaxy candidates that fell within the localisation region. We show the full list of host galaxy candidates and their \textsc{path} prior and posterior probabilities in Tab.~\ref{tab:pathanalysis}. We assumed that the host was detected in the deep imaging data. The host galaxy association analysis is inconclusive as there are four galaxies with posterior probabilities greater than 10 per cent. The candidate near the western edge of the detection beam footprint and marked as galaxy 1 in Fig.~\ref{fig:localisations} has the highest probability of about 17 per cent. It is spatially coincident with the infrared source WISEA~J042944.01$-$273023.2. Galaxy 2 is located near the northwestern border of the localisation region and has only a slightly lower association probability of about 15 per cent. Galaxy 3 has a slightly smaller angular extent, is located in the northwestern corner of the localisation, and has a posterior probability of 13 per cent.

The FRB is unresolved below the DM smearing of our data. It exhibits marginally significant scatter-broadening at the 1 to 2-$\sigma$ level as estimated from the marginalised $\tau_s$ posterior throughout the band and in the band-integrated profile, as shown in Fig.~\ref{fig:scattering}. Interestingly, the $\tau_s$ posterior peaks at non-zero values; see the bottom right panel in Fig.~\ref{fig:scattering}, suggesting a small but genuine scattering contribution. This is different from the other two FRBs whose $\tau_s$ posterior samples pile up at zero. However, the $\tau_s$ scaling with frequency is approximately constant, and its values are significantly below the DM smearing times. We conclude that its pulse width is primarily determined by instrumental DM smearing.

\subsubsection{FRB~20210202D}

We found FRB~20210202D in data obtained on 2021-02-02 UTC commensally with the MeerTime LSP \citep{2020Bailes} in an observation of the Galactic millisecond pulsar (MSP) PSR~J1946$-$5403 as part of its MSP pulsar timing sub-project. The FRB has a S/N of 18.6, a S/N-optimising DM of $609.16~\text{pc} \: \text{cm}^{-3}$, an extragalactic DM of $486~\text{pc} \: \text{cm}^{-3}$, a $\text{W}_\text{50p}$ pulse width of 2.3~ms, an inferred peak flux density $> 169~\text{mJy}$, and a fluence $> 0.4~\text{Jy}~\text{ms}$. We expect a host galaxy redshift range of [0.219, 0.516] or [0.281, 0.764] when considering the uncertainty in the DM -- redshift relation. Given the southern declination of the FRB field of about $-54$~deg, it is outside the observing regions of deeper wide-field optical surveys such as PS1, and very sparse optical data are available. In Fig.~\ref{fig:localisations} we show the SkyMapper DR2 \textit{i}-band data \citep{2019Onken} with an approximate limiting magnitude of 21. We selected the galaxies within the localisation region from the SkyMapper database based on their \texttt{class star} classification and enforced strict data quality requirements. Out of those, we chose only the objects that were clearly extended by applying a cut in their PSF magnitude compared with their Kron aperture magnitude \citep{1980Kron}, which is a standard selection technique. We also filtered out candidates with Gaia DR2 \citep{2018Brown} star classifications and measured parallaxes. The apparent magnitudes have not been corrected for dust extinction. There are only three galaxies visible within the localisation region. The \textsc{path} analysis is inconclusive, as all three galaxies have comparable association probabilities, see Tab.~\ref{tab:pathanalysis}. The total probability for an unseen host is about 45~per cent, given the shallow optical coverage. The brightest galaxy in the eastern corner of the localisation region, labelled as galaxy 1, is marginally favoured with a probability near 25~per cent. If we assume that all host galaxy candidates were detected, the posterior probabilities increase to 44, 31, and 25 per cent for galaxies 1 through 3.

The FRB is unresolved below the intra-channel dispersive smearing of our data. It shows scattering times that are consistent with zero. Out of the MeerTRAP FRBs considered here, it is exceptionally narrow with a post-scattering pulse width $\text{W}_\text{50p}$ of 2.3~ms at 1.284~GHz.

\subsubsection{FRB~20210408H}

\begin{figure}
  \centering
  \includegraphics[width=0.49\textwidth]{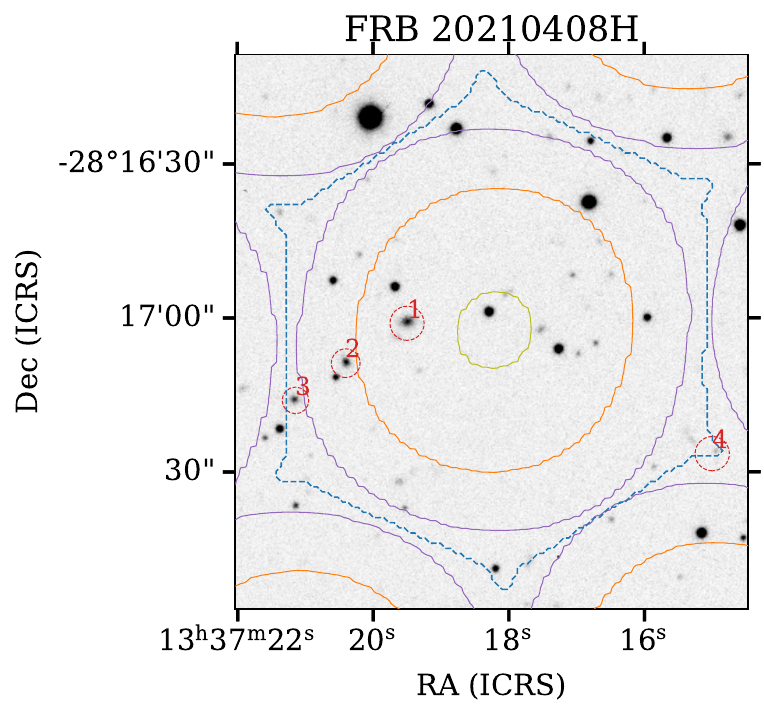}
  \caption{A deeper DECam \textit{z}-band image of the FRB~20210408H field that shows its host galaxy candidates more clearly than the Pan-STARRS1 data. The lines denote the same as in Fig.~\ref{fig:localisations}. Galaxy 1 is the brightest and most extended galaxy in that image.}
 \label{fig:decam}
\end{figure}

We discovered FRB~20210408H in data taken on 2021-04-08 UTC of the ESO444$-$G084 field observed commensally with the MHONGOOSE LSP. The FRB has a S/N of 14.7, a S/N-optimising DM of $1195.95~\text{pc} \: \text{cm}^{-3}$, an extragalactic DM of $1098~\text{pc} \: \text{cm}^{-3}$, a $\text{W}_\text{50p}$ pulse width of 6.4~ms, an inferred peak flux density $> 63~\text{mJy}$, and a fluence $> 0.4~\text{Jy}~\text{ms}$. The expected host galaxy redshift range is [0.886, 1.184] or [0.683, 1.608]. In Fig.~\ref{fig:localisations} we show existing Pan-STARRS1 DR1 \citep{2016Chambers} optical imaging data of the FRB field. The approximate limiting magnitudes are 23.4 (\textit{g}) and 22.7 (\textit{i}-band). For the \textsc{path} analysis, we selected the objects from the PS1 DR2 stacked object catalogue around the FRB's localisation that were clearly extended by applying a cut in PSF magnitude versus Kron magnitude. Strict data quality requirements were applied too. We used the centroid positions, the apparent \textit{i}-band magnitudes corrected for dust extinction \citep{1998Schlegel} and Kron radii as input for the \textsc{path} software. Only four host galaxy candidates within the localisation region fulfilled our selection criteria, see Tab.~\ref{tab:pathanalysis}. Galaxy 1 has a posterior probability of about 35~per cent. It is spatially coincident with the IR and UV sources WISEA~J133719.51$-$281700.5 and GALEXMSC~J133719.58$-$281700.9. Galaxy 2 is close to the sources WISEA~J133720.44$-$281708.8 and GALEXMSC~J133720.51$-$281708.8 and has an association probability near 17~per cent. Galaxy 3 coincides with the IR source WISEA J133721.17$-$281716.1 and has a probability of 9~per cent. Finally, galaxy 4 is located in the far southwest corner of the localisation regions and has a negligible posterior probability of about 4~per cent. While the association probabilities are not drastically different, galaxy 1 seems preferred overall. However, the probability of an unseen host is about 34 per cent. If we assume that all host galaxy candidates were detected, the posterior probabilities increase to about 53, 26, 14, and 6 per cent for galaxies 1 through 4. That is, galaxy 1 accounts for the majority of the posterior probability.

In Fig.~\ref{fig:decam}, we show additional and significantly deeper DECam imaging data of the FRB~20210408H field obtained in the \textit{z}-band filter, which is well suited for higher redshift objects. The data nicely show the extents of, and provide glimpses at the morphologies of, the candidate galaxies. Galaxy 1 is clearly the brightest and most extended galaxy within the localisation region, which further strengthens our conclusion from the \textsc{path} analysis that it is the favoured host. Interestingly, galaxy 4 appears only faintly in the \text{z}-band image. In the PS1 data, it was only detected in the \textit{i} and \textit{z}-band filters, where it is significantly brighter in the bluer \textit{i}-band wavelength range ($\sim$21.5 versus 23.2~mag). Neglecting the precise filter responses and remembering that these are Kron magnitudes (i.e.\ from unforced photometry), this might suggest that it is a lower redshift object and thereby disqualifies it as a host candidate. Galaxies 1 and 2 both increase in brightness from \textit{g} to \textit{z}-band, with a slight fall-off at \textit{y}, as expected for higher redshift objects. Similarly, galaxy 3 is brighter in \textit{z} than \text{i}-band, and again fainter in \textit{y}.

Unfortunately, all four galaxies lack redshift estimates in the literature. We used the data-driven local linear regression technique in a 5D magnitude and colour space developed by \citet{2016Beck} for the Sloan Digital Sky Survey DR12 and applied to PS1 DR2 by \citet{2020Tarrio} to estimate photometric redshifts for the host galaxy candidates. In particular, we used the PS1 DR2 stack photometry data for the galaxy candidates, i.e.\ the \textit{r}-band Kron magnitude and the four Kron colours (\textit{g} -- \textit{r}, \textit{r} -- \textit{i}, \textit{i} -- \textit{z}, \textit{z} -- \textit{y}), as input for the software and the training data set\footnote{\url{https://www.galaxyclusterdb.eu/m2c/relatedprojects/photozPS1}} provided by \citet{2020Tarrio}. As shown by those authors, using the Kron colours instead of the aperture colours results in essentially the same redshift estimates. Using this technique, we estimated photometric redshifts of the two brightest galaxies. Galaxy 1 (PS1 ID 74052043311949899) has a $z_\text{phot} = 0.45 \pm 0.08$, while galaxy 2 (ID 74052043349637297) has a $z_\text{phot} = 0.51 \pm 0.14$. Both estimates are based on all five features and local linear interpolation in the 5D feature space, i.e.\ quite robust. As \citet{2020Tarrio} investigated, the inferred photometric redshifts recover the spectroscopic measurements quite well in the range $0.1 < z_\text{spec} < 0.6$. For higher-redshift galaxies $z_\text{spec} > 0.6$, the technique seems to underestimate the redshift by up to $\sim$0.2 in the median. Redshift estimates for the other two galaxies (IDs 74052043381434904 and 74052043125301471) were unsuccessful, as they had two or more Kron magnitudes or features missing. At first glance, the galaxy redshifts seem slightly too low to reconcile with the FRB's observed DM of nearly $1196~\text{pc}~\text{cm}^{-3}$ and our expected host galaxy redshift ranges discussed above. This could suggest that they are unrelated foreground galaxies and that the actual FRB host galaxy is not visible in the PS1 imaging data. This is in line with our probability analysis for an unseen host. If the FRB indeed originated at $z \sim 1$ or above, we might need vastly deeper optical observations to detect its host. On the other hand, a more significant host galaxy DM contribution, together with the uncertainty in the cosmic DM -- redshift relation and any systematic error in the photometric redshift estimates, can account for the discrepancy. To illustrate the point, one only needs to moderately increase the combined Milky Way halo and host galaxy DM contribution to $\sim$200~$\text{pc}~\text{cm}^{-3}$, i.e.\ $\sim$150~$\text{pc}~\text{cm}^{-3}$ of host DM, to make galaxy 1's redshift estimate formally compatible with the expected redshift range at the 2-$\sigma$ level. The tension reduces further for increasing host contributions or if galaxy 2 is considered.

When looking at the FRB host galaxy database\footnote{\url{https://frbhosts.org/} -- Now defunct as of 2023-02-08.} and primarily focusing on the observed DM and host galaxy redshift, the highest-DM burst, FRB~20190614D, with a DM of $959.2~\text{pc}~\text{cm}^{-3}$, two plausible hosts at $z_\text{phot} \simeq 0.6$, and a host DM contribution $\sim$50~$\text{pc}~\text{cm}^{-3}$ \citep{2020Law} seems to be the closest match. The faintness of the galaxies (23 - 24~mag) appears to point to the first scenario discussed above, i.e.\ that the FRB's actual host is not visible in our current images. On the contrary, one could imagine FRB~20210408H to be a slightly closer variation of it, but with the difference in observed DM, $\sim$237~$\text{pc}~\text{cm}^{-3}$ ($\sim$344~$\text{pc}~\text{cm}^{-3}$ in the host's frame), coming mainly from the host galaxy or ionised material close to the source. These plasmas might not impart significant scattering on the FRB signal due to their proximity or the particular viewing geometry, in agreement with our data. Similarly, the DM smearing of our data could mask any lower-level amounts of scatter broadening.

The FRB is unresolved in our data below the intra-channel dispersion smearing and its scattering times are consistent with zero. Aside from the DM smearing, the FRB shows hints of being double-peaked, which is visible in both its dynamic spectrum and pulse profile.

\subsection{A post-cursor burst detection for FRB~20210202D}
\label{sec:postcursorburst}

\begin{figure}
  \centering
  \includegraphics[width=\columnwidth]{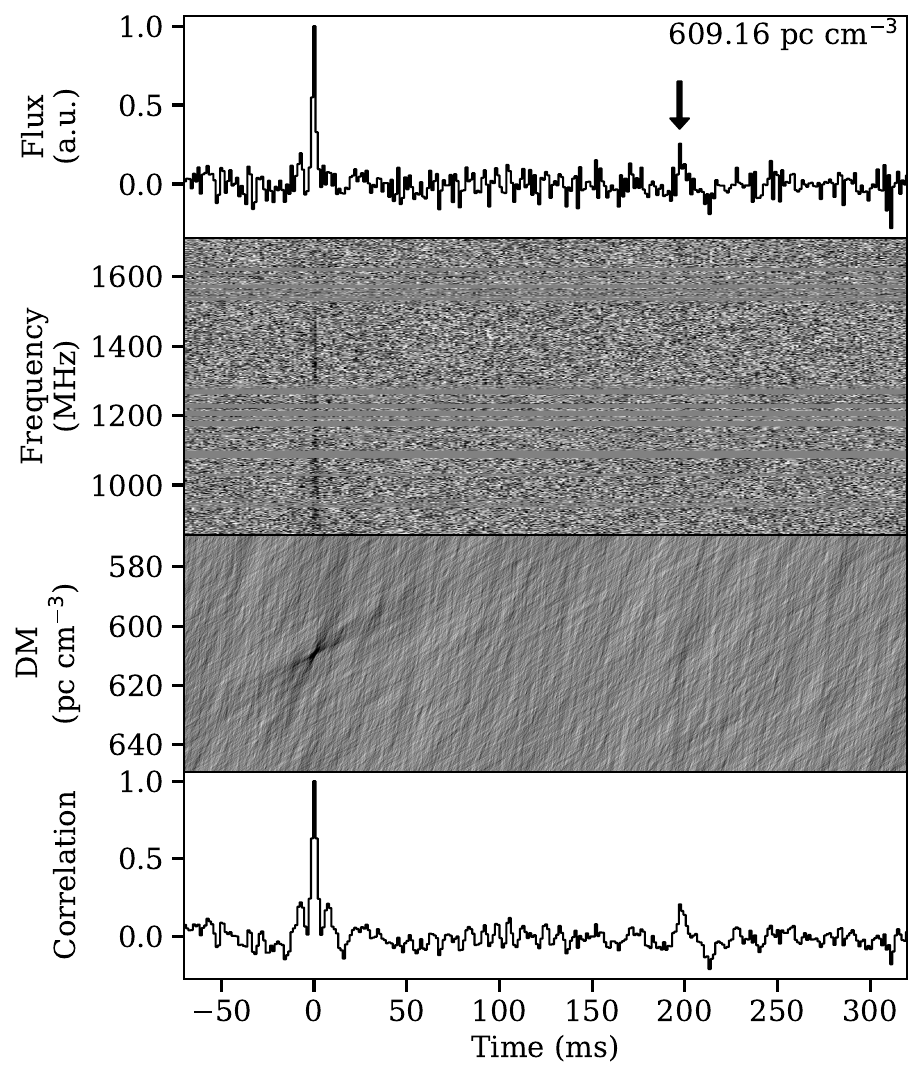}
  \caption{Post-cursor burst detection of FRB~20210202D. In the panels, we show from top to bottom: the dedispersed pulse profiles, a dedispersed dynamic spectrum, a trial DM versus time plot, and the cross-correlation power from correlating a narrow section of profile data centred on the main FRB component with the dedispersed time series.  The data are displayed at their native frequency resolution, but we averaged every four time samples for clarity. The post-cursor pulse or secondary emission component is faintly visible in all panels around 200~ms after the main burst and appears to have a comparable DM.}
 \label{fig:postcursorburst}
\end{figure}

As shown in Fig.~\ref{fig:postcursorburst}, FRB~20210202D seems to be followed by a significantly weaker repeat pulse or secondary emission component about 200~ms after the main pulse envelope. This is interesting as it could indicate that the FRB is a repeater. While the primary burst has a S/N of 18.6, the post-cursor is significantly fainter with an approximate S/N of 5.7. As such, it would fail our S/N discovery threshold as an individual burst. Its separation is about $644 \pm 8$ time samples, or $197 \pm 3~\text{ms}$, with the uncertainty coming from the sample averaging and its pulse width. As shown in the trial DM versus time plot, the post-cursor seems to have a comparable DM to the main burst, providing additional support that it might indeed be emission from the same source. Although faint, it can be seen across $\sim$730~MHz of bandwidth. While the main burst seems to become fainter with increasing frequency, the post-cursor appears to do the opposite, i.e.\ it might have a flatter spectral index than the primary burst. This agrees well with the fact that pulses from repeating FRBs show widely varying spectral indices \citep{2016Spitler}, although the bursts reported here have broadband spectra and show no frequency down-drift. However, we must caution that while relative spectral index comparisons are appropriate, the bandpass is not calibrated on an absolute scale. We also have to note that periodic zero-DM RFI was present in the data before excision, which, although very unlikely, could still potentially affect the underlying statistics of the data. The bottom panel of Fig.~\ref{fig:postcursorburst} shows the cross-correlation power from correlating a narrow section of data around the dedispersed main burst about 34 bins or 136 time samples wide (our ``template'') with the dedispersed time series. The correlation power exceeds the noise floor visibly at the post-cursor location. Finally, the post-cursor profile appears to be somewhat wider than the main burst and of approximately constant width across the band. This would mean that while the main burst is intrinsically narrower than the intra-channel dispersion smearing of our data for that DM ($\sim$4.2~ms at 1~GHz), the post-cursor's intrinsic width must exceed that.

In the following, we estimated whether the post-cursor events occurred simply because of random chance coincidence due to baseline noise fluctuations. Assuming typical values of $\pm 2$ DM trials centred on the S/N-optimising best-determined FRB DMs (the S/N versus trial DM curves are well peaked and fall off steeply), a search window of $\pm 200$~ms either side of the main burst, an average post-cursor width of 8 time samples, and a sample size of 11 MeerTRAP FRBs considered, the total number of trials is $N_\text{t} = 11 \times 5/8 \times 400~\text{ms} / 306.24~\mu \text{s} \approx 8980$. Neglecting RFI and assuming normally distributed radiometer noise with zero mean $\mu$ and unit standard deviation $\sigma$, the tail probability of detecting a $\text{S/N}_\text{p} = 5.7$ event is $P_\text{samp} ( \text{S/N}_\text{p} ) = \text{SF} \left[ \mathcal{N} ( \mu, \sigma^2; \text{S/N}_\text{p} ) \right] \approx 6 \times 10^{-9}$, where $\text{SF} = 1 - \text{CDF}$ denotes the survival function of the standard normal distribution $\mathcal{N}$. As expected, the probability is extremely low. Considering $N_\text{t}$ fully independent samples, the total probability is $P_\text{tot} ( \text{S/N}_\text{p} ) = 1 - (1 - P_\text{samp})^{N_\text{t}} \approx 5.4 \times 10^{-5}$, i.e.\ the significance of the post-cursor detection reduces from 5.7 to 3.9-$\sigma$ when accounting for the number of trials, which is still reasonably high. The significance of an 8 S/N burst reduces to 6.8-$\sigma$ for the same parameters.

\subsection{MeerTRAP survey performance and completeness}
\label{sec:surveycompleteness}

\begin{figure}
  \centering
  \includegraphics[width=\columnwidth]{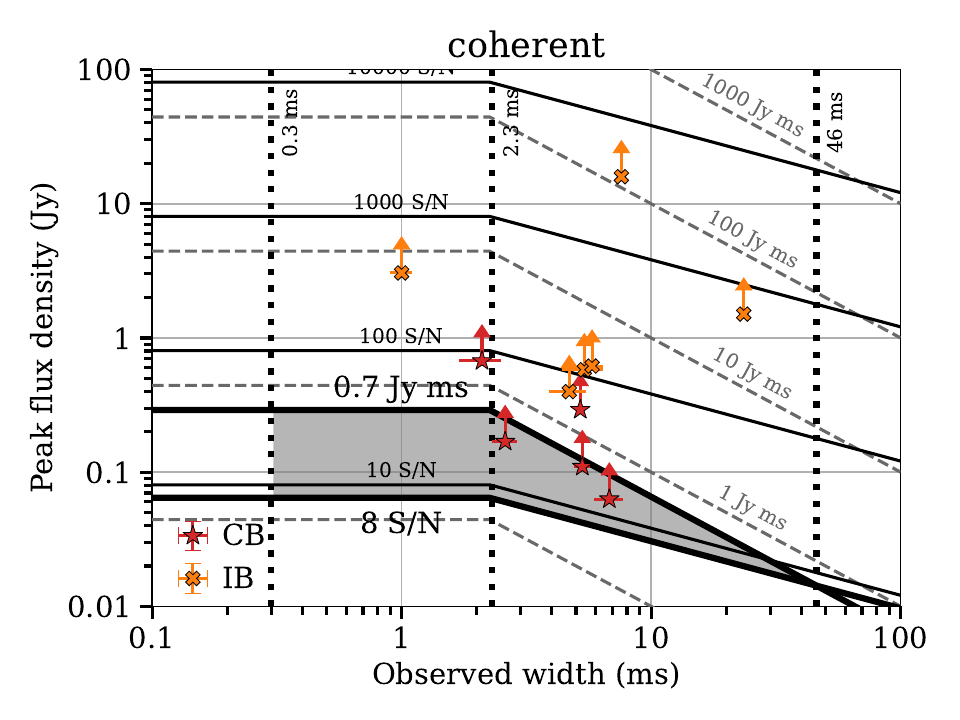}
  \includegraphics[width=\columnwidth]{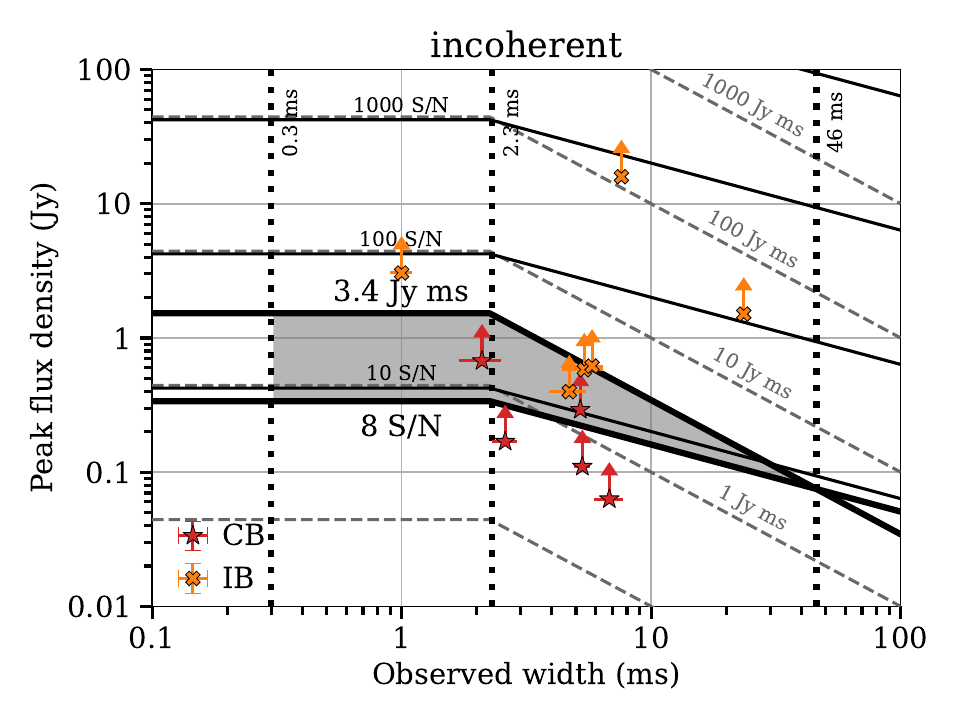}
  \caption{Survey performance curves and inferred fluence completeness thresholds of the coherent (top) and incoherent (bottom) MeerTRAP transient surveys at L-band. We show S/N curves $\propto \text{W}_\text{obs}^{-0.5}$ as solid black lines, while fluence curves $\propto \text{W}_\text{obs}^{-1}$ have dashed gray lines. We highlight the 8.0 S/N and the fluence completeness threshold curves with thick solid black lines. The shaded areas are the fluence incompleteness regions. The surveys become successively more complete for bursts located above the topmost thick fluence completeness lines. We marked the parameters of the MeerTRAP FRBs presented here, those published and under review, separated by discovery beam type for reference. Note that their inferred peak flux densities are based on offline refined data with S/N values and equivalent widths that are typically higher than from their real-time discoveries. The curves have been drawn flat below 2.3~ms.}
 \label{fig:fluencecompleteness}
\end{figure}

Crucial survey parameters of the MeerTRAP transient surveys are their limiting peak flux densities and fluences, and their fluence completeness thresholds $F_\text{c}$ \citep{2015Keane, 2019James}, which have not been systematically estimated before. We derived them in the following. We performed the vast majority of the surveys with a detection threshold $\text{S/N}_\text{th} = 8.0$ for the single-pulse pipeline. We estimated the performance parameters for each survey based on a modified version of the single-pulse radiometer equation \citep{1985Dewey}
\begin{equation}
    S_\text{peak} \left( \text{S/N}, \text{W}_\text{eq}, \vec{a} \right) = \text{S/N} \: \beta \: \eta_\text{b} \: \frac{ T_\text{sys} + T_\text{sky} }{ G \sqrt{b_\text{eff} N_\text{p} \text{W}_\text{eq}} } \: a_\text{CB}^{-1} \: a_\text{IB}^{-1},
    \label{eq:radiometer}
\end{equation}
where $S_\text{peak}$ is the peak flux density, $\vec{a}$ is the parameter vector, $\beta$ is the digitisation loss factor, $\eta_\text{b}$ is the beam-forming efficiency, $G$ is the telescope forward gain, $b_\text{eff}$ is the effective bandwidth, $N_\text{p}$ is the number of polarisations summed, $\text{W}_\text{eq}$ is the observed equivalent boxcar pulse width, $T_\text{sys}$ and $T_\text{sky}$ are the system and sky temperatures, and $a_\text{CB}$ and $a_\text{IB}$ are the attenuation factors of the detection CB and the IB. The overall performance parameters include a total telescope gain (64 antennas) of $\sim$2.77~$\text{K}~\text{Jy}^{-1}$ \citep{2020Bailes}, $N_\text{p} = 2$, a median system temperature across the band of 19~K including spill-over and atmospheric terms\footnote{See the measured system temperature over aperture efficiency data provided by the MeerKAT observatory team at \url{https://skaafrica.atlassian.net/wiki/spaces/ESDKB/pages/277315585/MeerKAT+specifications}},  a digitisation loss factor of essentially unity for our 8-bit sampled data \citep{2001Kouwenhoven}, and a beam-forming efficiency close to unity \citep{2021Chen}.

For our FRB discoveries, we used Eq.~\ref{eq:radiometer} with the offline refined measured values of S/N, $\text{W}_\text{eq}$, $b_\text{eff}$, $N_\text{ant,cb}$, $a_\text{IB}$, and $a_\text{CB} = 1$ given in Tab.~\ref{tab:burstproperties} to estimate their peak flux densities $S_\text{peak}$ and fluences $F = S_\text{peak} \: \text{W}_\text{eq}$. The sky temperature was fixed to the mean values at their position from the \citet{1982Haslam} all-sky atlas \citep{2015Remazeilles} scaled to 1284~MHz using a power law exponent of $-2.6$ \citep{1987Lawson}. While we have a good handle on $a_\text{IB}$ for each FRB (see Tab.~\ref{tab:burstproperties}), $a_\text{CB}$ is essentially unknown, as we lack information in which part of the CB response the FRBs occurred. To illustrate this, the primary beam correction factors $a_\text{CB}^{-1}$ amount to only about 2, 6, and 13~per cent for the FRBs in this paper, while the CB corrections could be significantly higher given the narrow Sinc function-like response of the array. That means that the FRB fluences could be severely underestimated by a factor of a few. When modelling the MeerTRAP survey performance, we used Eq.~\ref{eq:radiometer} with an effective bandwidth presented to the real-time single-pulse search software of $b_\text{eff} = 540~\text{MHz}$, an observed boxcar equivalent width
\begin{equation}
    \text{W}_\text{eq} = \sqrt{\text{W}_\text{i}^2 + t_\text{s}^2 + t_\text{dm}^2 + \text{W}_\text{scat}^2},
    \label{eq:observedwidth}
\end{equation}
 that includes contributions from the sampling time $t_\text{s}$, and typical values for the intra-channel dispersive smearing $t_\text{dm}$ and scattering $\text{W}_\text{scat}$, a median sky temperature of $2.73~\text{K}$, and an average of 40 MeerKAT antennas contributing to the coherent and 58 to the incoherent survey. We assumed the latter value because that is the number of antennas guaranteed to be available as stated by the observatory team. We defer a spatially-resolved performance analysis to future work.

In its current configuration, the MeerTRAP surveys are limited by the broad channelisation of the data and the accompanying intra-channel dispersion smearing, which particularly affects intrinsically narrow high-DM FRBs. For instance, assuming a nominal FRB with a DM of $1000 \: \text{pc} \: \text{cm}^{-3}$ and a channel bandwidth of $b_\text{c} \approx 0.836~\text{MHz}$, the DM smearing $t_\text{dm}$ at 1.284~GHz is 3.28~ms and increases to 11.08~ms at the bottom of the band, according to Eq.~\ref{eq:dmsmearing}. Together with the finite sampling time $t_\text{s} = 306.24~\mu \text{s}$, the minimum resolvable width is (Eq.~\ref{eq:observedwidth}), therefore, $\text{W}_\text{eq}^\text{min} = \sqrt{t_\text{s}^2 + t_\text{dm}^2} \approx 3.3~\text{ms}$ at the centre of the band.

For the typical observing setup described above, the limiting peak flux densities are about 60 and 340~mJy (150 and 770 mJy~ms fluence) for a S/N = 8 1~ms burst smeared to an observed width of $\sim$2.3~ms ($t_\text{dm} = 2~\text{ms}$, $\text{W}_\text{scat} = 1~\text{ms}$) at boresight and the CB centre in the coherent and incoherent surveys, respectively. For the best case that all 64 antennas are available and that there is no RFI, i.e.\ that all 770~MHz of on-sky bandwidth can be used, the limiting peak flux densities decrease to about 50 and 270~mJy (120 and 610 mJy~ms fluence).

Based on the modified radiometer equation (Eq.~\ref{eq:radiometer}), we estimated the MeerTRAP fluence completeness thresholds $F_\text{c}$ following the prescription given by \citet{2015Keane}. Namely, an idealised boxcar-shaped burst of observed equivalent width $\text{W}_\text{eq}$ and given S/N has a fluence $F = S_\text{peak} \: \text{W}_\text{eq}$, with $S_\text{peak}$ as defined in Eq.~\ref{eq:radiometer} and the values of the parameter vector $\vec{a}$ as discussed above. $F_\text{c}$ is then determined from the widest confidently-detected burst of width $\text{W}_\text{eq}^{\star} = \text{max} \left( \text{W}_{\text{eq},i} \right)$ as
\begin{equation}
    F_\text{c} = S_\text{peak} \left( \text{S/N}_\text{th}, \text{W}_\text{eq}^{\star}, \vec{a} \right) \: \text{W}_\text{eq}^{\star},
    \label{eq:fcut}
\end{equation}
where $\text{S/N}_\text{th} = 8.0$ is the threshold S/N of the surveys. The method is robust, as its completeness estimate is based on empirical measurements of the telescope's and detection pipeline's performance on actual astrophysical bursts. It effectively places the widest detected burst at the S/N detection threshold. Fig.~\ref{fig:fluencecompleteness} shows the resulting ``triangle'' fluence completeness plots for both MeerTRAP transient surveys. The observed burst width ranges from our sampling time to the maximum FRB pulse width observed so far, $\text{W}_\text{eq}^{\star} \approx 46~\text{ms}$. We show only the pulse width range up to 100~ms for clarity, although we typically consider candidates up to $\sim$300~ms in boxcar width. However, we sometimes had to discard the widest candidates beyond $\sim$100~ms, e.g.\ at times of strong RFI. However, exceptionally wide pulses are usually detected through their bright features at smaller widths within that search range anyway. The minimum observable burst width given by Eq.~\ref{eq:observedwidth} for an infinitesimally small intrinsic width and assuming typical values of $t_\text{dm} = 2~\text{ms}$ and $\text{W}_\text{scat} = 1~\text{ms}$, is $\text{W}_\text{eq}^\text{min} \approx 2.3~\text{ms}$. This choice is appropriate for the current sample of MeerTRAP FRBs, which is apparent from their positions in Fig.~\ref{fig:fluencecompleteness}. Namely, almost all are located near or beyond $\text{W}_\text{eq}^\text{min}$. An exception is FRB~20200915A \citep{2022Rajwade}, which was detected in 4096 frequency channel data, i.e.\ in data with a channel bandwidth four times smaller than the others ($\sim$0.209~MHz). Consequently, the lowest observable width is four times smaller, $\sim$0.6~ms at 1.284~GHz and its DM of $740.5~\text{pc} \: \text{cm}^{-3}$. Its observed boxcar equivalent width is $1.0 \pm 0.1$~ms, approximately half the pulse width given in \citet{2022Rajwade}. The difference is due to a lack of sample-accurate analysis tools like \textsc{scatfit} in the earlier work. In any case, bursts that are intrinsically narrower than $\text{W}_\text{eq}^\text{min}$ get smeared by the instrumentation to at least this width. Their measured S/N, inferred peak flux densities, and fluences are therefore underestimated, which is visible as the plateaus in the S/N and fluence curves in Fig.~\ref{fig:fluencecompleteness}.

Importantly, the inferred peak flux densities of the FRBs displayed in Fig.~\ref{fig:fluencecompleteness} were computed from the offline refined S/N and pulse width measurements and not from the discovery values reported by the real-time pipeline. Initially, the real-time S/N values were typically significantly below the refined ones. This discrepancy has since been rectified. The inferred fluence completeness thresholds are $0.66$ and $3.44~\text{Jy}~\text{ms}$ for the coherent and incoherent MeerTRAP surveys, respectively. As expected, the coherent survey is approximately five times more sensitive than the incoherent. The latter limit is comparable to the ones adopted at the Parkes \textit{Murriyang} radio telescope \citep{2015Keane, 2016Champion, 2018Bhandari}, while the former pushes into the phase space previously only accessible by the Arecibo Telescope \citep{2014Spitler} or FAST \citep{2021Niu}.

\subsection{Inferred FRB all-sky rates}
\label{sec:frballskyrate}

\begin{figure}
  \centering
  \includegraphics[width=\columnwidth]{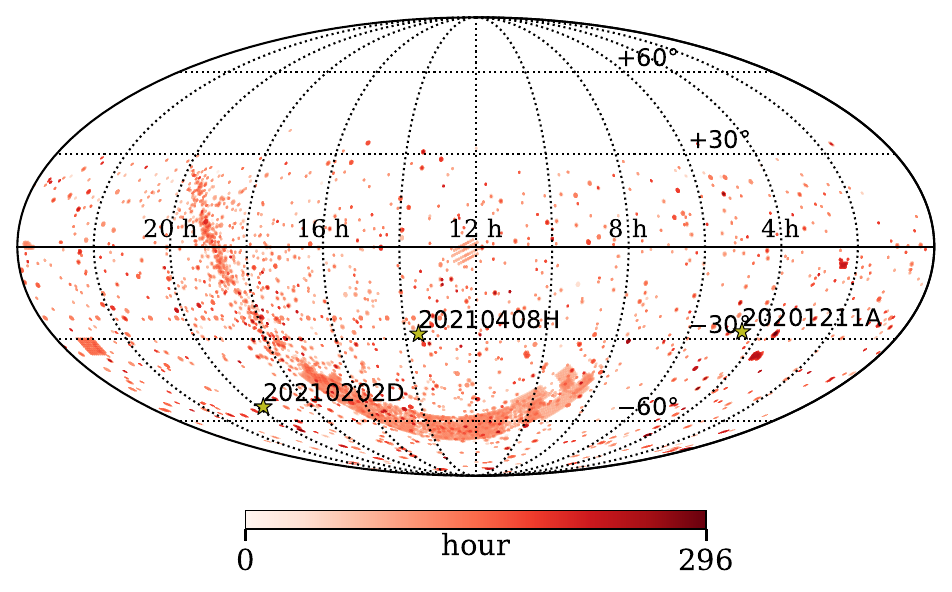}
  \includegraphics[width=\columnwidth]{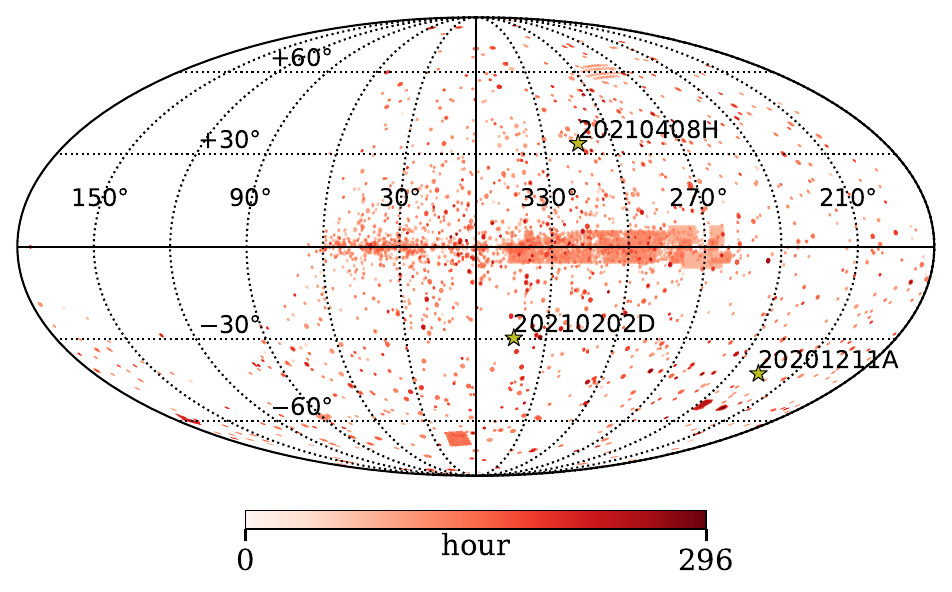}
  \caption{Mollweide projections in equatorial (top) and Galactic coordinates (bottom) of the sky coverage of the MeerTRAP incoherent survey at L-band from 2019 June until the end of 2021 December. We marked the locations of the FRB discoveries presented in this work.}
 \label{fig:surveycoverage}
\end{figure}

\begin{table*}
\caption{Parameters of the MeerTRAP transient surveys at L-band that are centred at 1284 MHz with 856 MHz of digitised and $\sim$770~MHz of on-sky bandwidth, of which typically $\sim$540~MHz are RFI-free. We present the survey coverages $c_s$, the fluence completeness limits $F_\text{c}$, the numbers of detected FRBs in the time frame covered in this work, and the inferred FRB all-sky rates, assuming a detection efficiency $\eta_p = 0.5$ of our single-pulse search pipeline.}
\label{tab:allskyrates}
\begin{tabular}{lcccccc}
\hline
Survey      & $t_\text{obs}$    & $\left< A_{0.5} \right>$    & $c_s$                         & $F_\text{c}$    & $N_\text{FRB} \left(> F_\text{c}\right)$    & $R_\text{FRB} \left( >F_\text{c}\right)$\\
            & (d)   & ($\text{deg}^2$)  & ($\text{deg}^2 \: \text{h}$)  & (Jy ms)           &                                               & ($10^3 \: \text{sky}^{-1} \: \text{d}^{-1}$)\\
\hline
Coherent                & 317.5 & 0.19  & 1448    & 0.66  & 6     & $8.2_{-4.6}^{+8.0}$\\
Incoherent (total)      & 317.5 & 0.97  & 6662    & 3.44  & 7     & $2.1_{-1.1}^{+1.8}$\\
Incoherent (subtracted) & 317.5 & 0.78  & 5944    & 3.44  & 5     & $1.7_{-1.0}^{+1.8}$\\
\hline
\end{tabular}
\end{table*}

We present a map of the sky coverage of the MeerTRAP surveys in Fig.~\ref{fig:surveycoverage}, which was generated using our \textsc{skymap} software\footnote{\url{https://github.com/fjankowsk/skymap/}}. The survey coverage $c_s$ is defined as
\begin{equation}
    c_s = \sum_i t_\text{obs,i} \: A_{0.5,i},
    \label{eq:surveycoverage}
\end{equation}
where the sum runs over all survey pointings $i$, $t_\text{obs,i}$ is the observing time, and $A_{0.5,i}$ is the covered half-power beam sky area in that pointing. For the incoherent beam search, $A_{0.5,i}$ is the half-power area of the MeerKAT primary beam, whereas, for the coherent beam search, it is the sum of the half-power areas of the typically up to 768 coherent tied-array beams that tile the primary beam. In particular, we used a model derived from astro-holographic measurements of the MeerKAT Stokes I primary beam response \citep{2021Asad, 2022DeVilliers} at half-power and evaluated at the centre of the observing band, see Fig.~\ref{fig:beamarea}. The model is consistent with a cosine-tapered field illumination pattern up to 2.5~deg radial distance from the beam centre \citep{2020Mauch}. For the tied-array beam search, we summed the areas of at least half-power of the total simulated beam tiling patterns on the sky \citep{2021Chen} for typical observing configurations with the CBs overlapping at 25~per cent maximum. As shown in Fig.~\ref{fig:beamarea}, the mean half-power areas $A_{0.5}$ at 1284~MHz and using a maximum of 768 CBs are approximately 0.97 and $0.19~\text{deg}^2$, respectively. This corresponds to a half-power area per individual CB of about $0.9~\text{arcmin}^2$. We verified those values by looking at the histograms of the CB sizes reported by the system sensors across the time frame of interest. Their medians match those numbers above. In other words, the area covered by the incoherent beam at half-power is roughly $5$ times that of the total CB pattern. The CB $A_{0.5}$ values are more variable, as they depend on the number of beams searched, the specific antennas used for beam-forming (their maximum baseline), and the projected foreshortening of the array with increasing hour angle. Periods during which we know that the detection performance of the MeerTRAP pipeline was significantly reduced, e.g.\ because of known software issues, have been excluded from the survey coverage estimates. The survey coverages between 2019 June and the end of 2021 December are listed in Tab.~\ref{tab:allskyrates}. The total time on sky amounted to 317.5~d during that period, equating to an average of 8.1 observing hours per day.

Based on the survey coverage, we then estimated the FRB all-sky rate $R_\text{FRB}$ above the fluence completeness threshold $F_\text{c}$, assuming an isotropic FRB distribution on the sky, as
\begin{equation}
    \begin{split}
    R_\text{FRB} \left(>F_\text{c} \right) & = \frac{ N_\text{FRB} \left(>F_\text{c} \right) }{ \eta_p \: c_s } \\
    & = N_\text{FRB} \left(> F_\text{c} \right) \: \frac{ 24~\text{h} \: \text{d}^{-1} \: 41253~\text{deg}^2 \: \text{sky}^{-1} }{ \eta_p \: c_s \left[ \text{deg}^2 \: \text{h} \right] },
    \end{split}
    \label{eq:allskyrate}
\end{equation}
where $N_\text{FRB}$ is the number of detected FRBs above the threshold, $c_s$ is the survey coverage, and $0 < \eta_p < 1$ is the efficiency of the detection pipeline. Namely, $\eta_p$ is the efficiency (or probability) with which an FRB that is present in the digitised data stream is discovered after running the full detection pipeline chain. In our case it includes contributions from the employed RFI excision methods, the single-pulse search software, candidate clustering and sifting steps, known-source matching and multi-beam clustering, the machine-learning classifier, and human candidate vetting. These factors likely interact in complex ways, and $\eta_p$ is therefore challenging to quantify reliably. We conservatively assumed $\eta_p = 0.5$ and refer a systematic estimation to future work. Rigorous tests of the real-time pipeline with mock FRBs injected into the signal chain, as for instance pioneered at UTMOST \citep{2021Gupta} or CHIME \citep{2021CHIMECatalogue}, are needed to quantify its detection efficiency, assess its biases, and determine the survey selection function.

For our analysis, we considered the entire MeerTRAP L-band FRB sample discovered up to the end of 2021, i.e.\ those already published \citep{2022Rajwade}, the ones presented in this work, and those currently in preparation (e.g.~\citealt{2023Driessen, 2023Caleb}). When FRBs were discovered simultaneously in the IB and in one or several CBs, we included them in both the coherent and incoherent FRB samples, thereby double-counting them. In total, we based our analysis on 11 MeerTRAP FRBs, 6 CB and 7 IB detections, two of which were detected in both the IB and CBs. We assigned them to both samples to maximise the number of FRB detections in the low-number regime that we are currently in. For 6 and 7 discoveries we are fully dominated by the statistical error from the Poisson counting process; the 95~per cent confidence level relative errors are (56, 97) and (53, 88)~per cent for the low and high error bar, respectively. Assuming that systematic errors are present at the $\sim$25~per cent level \citep{2021CHIMECatalogue}, we need at least $\sim$54 FRB discoveries to reduce the counting error to a similar level.

Additionally, we accounted for a correlation between the FRB samples and therefore all-sky rates by excluding the two double-counted FRBs from the incoherent sample and reducing the IB sky area by that covered by the total CB grid at half-power. This ``subtracted'' survey therefore only includes the FRBs discovered in the IB that were not detected in the CBs and covers the sky area outside the central CB grid. Hence, it is more distinct from the coherent survey.

We show the survey parameters and inferred FRB all-sky rates in Tab.~\ref{tab:allskyrates}, where we quote them at the 95~per cent Poisson confidence level \citep{1986Gehrels}. Specifically, the derived rates are $8.2_{-4.6}^{+8.0}$, $2.1_{-1.1}^{+1.8}$, and $1.7_{-1.0}^{+1.8} \times 10^3 \: \text{sky}^{-1} \: \text{d}^{-1}$ for the coherent, incoherent (total) and incoherent (subtracted) surveys, respectively. The rates for the incoherent (total) and incoherent (subtracted) surveys are identical within the errors. The total instrumental MeerTRAP detection rate or survey yield, irrespective of the discovery beam type, is 11 FRBs per 317.5~d of on-sky time, or approximately one FRB discovery every $\sim$29~d on average for the current sample.

\subsection{FRB population parameter estimates}
\label{sec:frbpopulationparameters}

\begin{figure}
  \centering
  \includegraphics[width=\columnwidth]{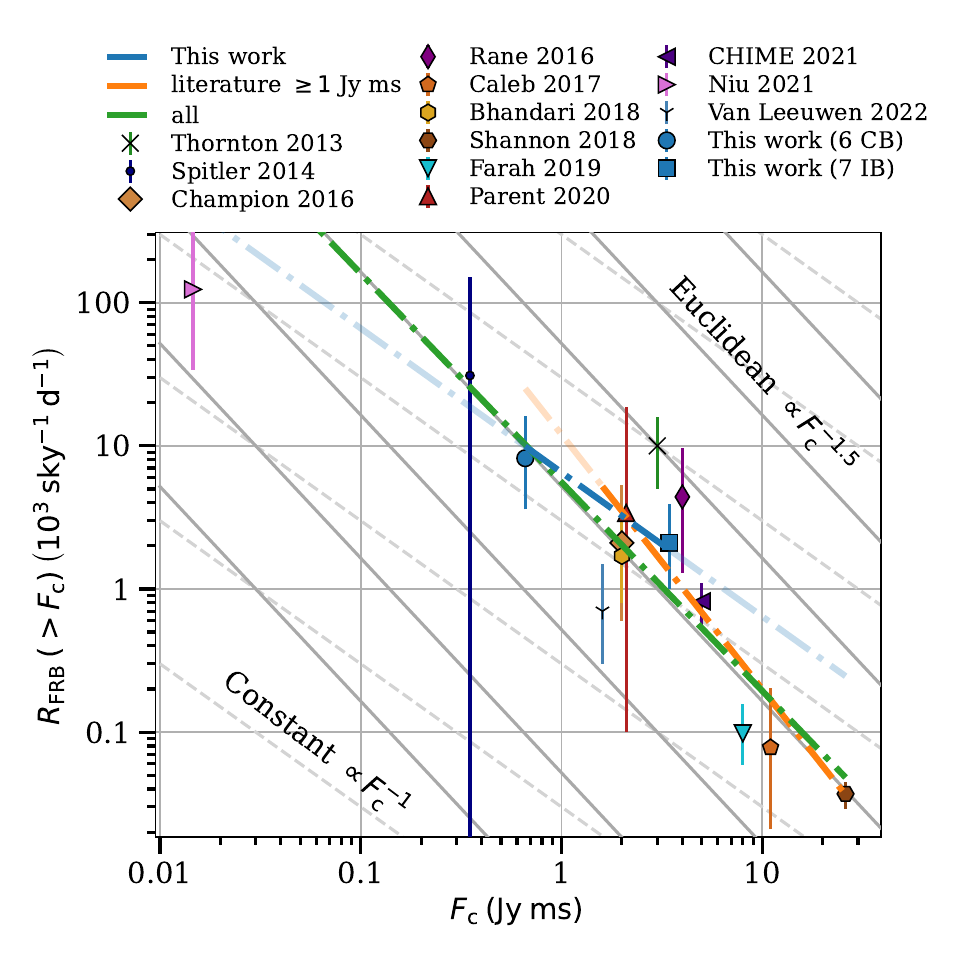}
  \includegraphics[width=\columnwidth]{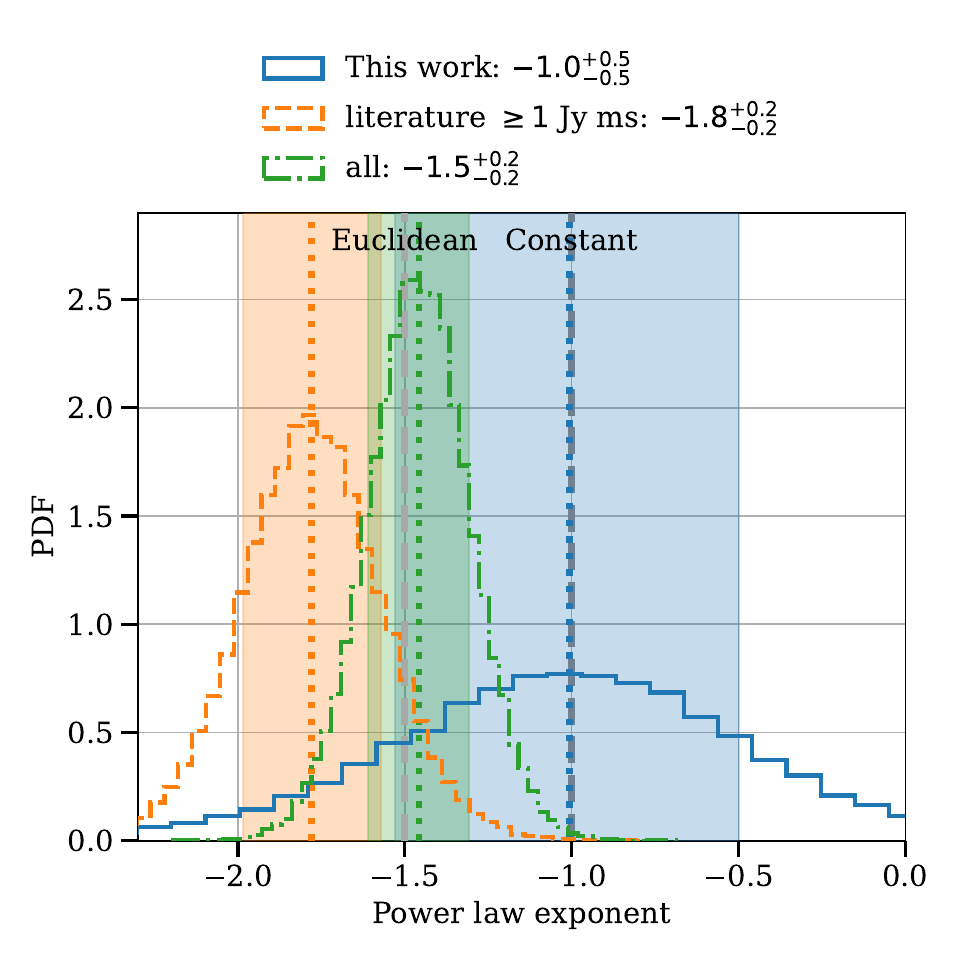}
  \caption{Inferred FRB all-sky rates. Top: We show the FRB all-sky rates inferred from the MeerTRAP surveys at L-band as a function of fluence completeness threshold $F_\text{c}$ compared with a selection of rates from the literature. The literature rates are from observations at various radio frequencies. We present the best-fitting power law functions to our rates, the high-fluence literature rates $\geq 1$~Jy~ms, and the combined data set. Bottom: Histograms of the posterior samples of the power law exponents from the fits in the top panel where we shaded the 68~per cent credibility ranges. The power law scaling between the MeerTRAP all-sky rates is consistent with a constant and Euclidean scaling, but it is appreciably flatter than that of the literature sample obtained at higher fluences. There seems to be a break or turn-over in the FRB all-sky rate versus limiting fluence relation somewhere below 1 - 2~Jy~ms. This could mean that the MeerTRAP FRB sample already probes the transition region from the local to the more distant Universe.}
 \label{fig:allskyrate}
\end{figure}

\begin{figure}
  \centering
  \includegraphics[width=\columnwidth]{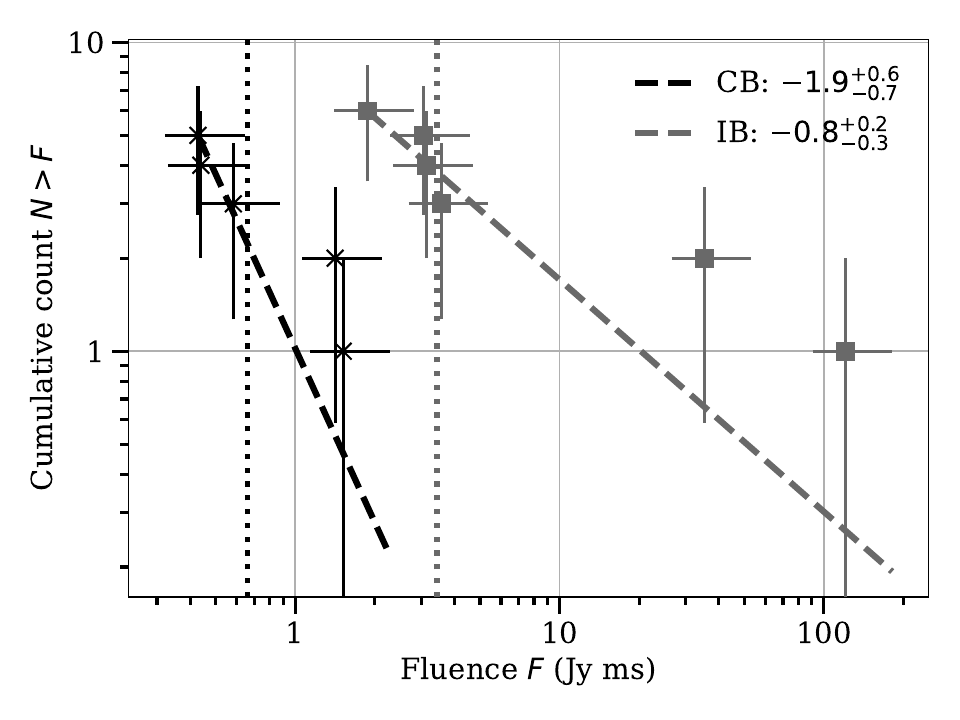}
  \caption{Cumulative source counts ($\log N - \log F$) for the MeerTRAP coherent and incoherent surveys. We show the observed counts with the best-fitting Pareto distributions overlaid. The scaling exponents $\delta = -\alpha$ were determined using an unbinned likelihood method; cumulative counts are displayed for convenience. The CB discoveries appear consistent with the Euclidean scaling within a survey. The IB count distribution is somewhat flatter, but agrees with Euclidean if the brightest IB FRB is excluded.}
 \label{fig:sourcecounts}
\end{figure}

In Fig.~\ref{fig:allskyrate}, we compare the FRB all-sky rates from the MeerTRAP L-band surveys with a selection of rates from the literature, obtained at different telescopes, frequencies, fluence thresholds, and survey selection functions. The literature rates come from \citet{2013Thornton}, \citet{2014Spitler}, \citet{2016Champion}, \citet{2016Rane}, \citet{2017Caleb}, \citet{2018Bhandari}, \citet{2018Shannon}, \citet{2019Farah}, \citet{2020Parent}, \citet{2021CHIMECatalogue}, \citet{2021Niu}, and \citet{2022VanLeeuwen}. In our comparison, we assumed a flat spectral index for the FRB population, as the frequency dependence of the FRB emission is still highly uncertain. For instance, \citet{2019Macquart} showed that there is a large degree of spectral modulation in bright ASKAP FRBs discovered at 1.4~GHz with perhaps a mean spectral index $\alpha = -1.5_{-0.3}^{+0.2}$ ($F \propto \nu^\alpha$) similar to that of the Galactic pulsar population \citep{2018Jankowski}. However, the low number of discovered FRBs in surveys at 300-400~MHz \citep{2020Parent} or 843~MHz \citep{2019Farah} suggests a significantly flatter spectral index or a spectral turnover below 1~GHz. Hence, assuming $\alpha = 0$ is a standard and conservative approach \citep{2021CHIMECatalogue}.

We fit a power law of the form
\begin{equation}
    R_\text{FRB} \left( > F_\text{c} \right) = R_\text{FRB,0} \left( > F_\text{c,0} \right) \left( \frac{ F_\text{c} }{F_\text{c,0}} \right)^{a},
    \label{eq:ratescaling}
\end{equation}
where $R_\text{FRB,0}$ is the FRB all-sky rate at the reference fluence threshold $F_{c,0}$ and $a$ is the power law exponent, to the rate versus fluence threshold data. We used the \textsc{pymc} Bayesian modelling and Markov chain Monte Carlo software suite \citep{2016Salvatier} in version 5.4, where we assumed mildly-informative Gaussian priors centred at $-1.5$ for the power law exponent and centred on the median rate in the data set for the normalisation. Additionally, we multiplied the rate uncertainties by a constant factor to account for error underestimation, on which we placed a lognormal prior centred at unity which was estimated during the sampling process. We separately fit the literature data $\geq 1$~Jy~ms (i.e.\ the rates from all surveys except the most sensitive ones by Arecibo and FAST), the inferred MeerTRAP all-sky rates from this work, and the entire data set. We show the best fits in the top panel of Fig.~\ref{fig:allskyrate} and histograms of the marginalised posteriors of the power law exponent in the bottom panel. For the entire data set, the correction factor for error underestimation has a median of $1.5_{-0.2}^{+0.2}$ with a tail towards higher values, as expected. As shown in Fig.~\ref{fig:allskyrate} bottom panel, the power law scaling between the MeerTRAP rates agrees well with an FRB constant space density scaling and is consistent with an Euclidean scaling within the errors. However, it is appreciably flatter than that of the literature rates above 1~Jy~ms. The difference in median power law exponent $a$ is significant at the 1.4-$\sigma$ level, where $\sigma$ is the quadrature sum of the uncertainties of $a$ from the regression, i.e.\ $\sigma = \sqrt{u_{a,\text{lit}}^2 + u_{a,\text{mk}}^2}$. Here, the uncertainty $u_{a,\text{mk}}$ on the MeerTRAP scaling exponent from the two-point estimate clearly dominates $\sigma$. If only $u_{a,\text{lit}}$ is considered, the significance becomes 4-$\sigma$. If we use the MeerTRAP IB (subtracted) rate instead in the fit, the best-fitting $a$ becomes $-1.1_{-0.6}^{+0.5}$ and the significance reduces slightly to 1.2-$\sigma$, but the overall result is the same. The MeerTRAP scaling is flatter than Euclidean and flatter than that of the literature measurements at higher fluences. As the number of MeerTRAP FRBs increases, the relative errors on the inferred rates decrease, and the power law exponent between them will become better constrained. Overall, there appears to be a break or turn-over in the FRB all-sky rate versus limiting fluence relation somewhere below 1 - 2~Jy~ms. The MeerTRAP scaling extrapolates near the FAST rate at the so far lowest limiting fluence, as shown by the slightly transparent blue line in Fig.~\ref{fig:allskyrate}.

The above analysis of the scaling of the inferred FRB all-sky rates with fluence completeness threshold provides an indirect or inter-survey measurement of the FRB population's fluence distribution. It is most suited for surveys with low numbers of detections, as the derived rate is an integral quantity across all the FRBs detected. A more direct and intra-survey approach is to look at the FRB source counts, i.e.\ their $\log N - \log F$ distributions. We display the cumulative source count distributions for the current MeerTRAP L-band sample of 11 FRBs in Fig.~\ref{fig:sourcecounts}, separated by survey. In this analysis, we assigned the two multi beam-type FRBs exclusively to the sample of their highest detection S/N and not also the other. Shown are the cumulative or integral counts above a limiting fluence. We assumed asymmetric fluence errors of 25 and 50~per cent on the best-determined values and Poissonian errors $\sqrt{N}$ on the counts $N$. All fluences were corrected by the attenuation of the FRBs in MeerKAT's primary beam response as in Eq.~\ref{eq:radiometer}. Where FRBs are well localised to either a single CB or by synthesis imaging, their $a_\text{IB}$ values are well established from the primary beam models. For more poorly localised (IB) bursts, $a_\text{IB}$ is the minimum attenuation (highest value) compatible with a non-detection in the central primary beam area tiled with CBs following the procedure in \citet{2022Rajwade}. For FRBs detected simultaneously in CBs and the IB, we used the combined beam information for their localisation and the IB data for their robust fluence estimates. Imaging-localised IB detections have the most reliable fluences, as their signals are only affected by the slowly-varying and well-characterised primary beam response. They completely avoid the more complex attenuation by the CB response and variations in beam-forming efficiency (array phasing). We employed an unbinned likelihood method to estimate the slopes of the integral source count distributions using \textsc{pymc}. In particular, we fit the empirical fluence distributions with a Pareto distribution whose CDF is
\begin{equation}
    C(x, x_\text{m}, \alpha) = 1 - \left( \frac{ x }{ x_\text{m} } \right)^{- \alpha} \propto \left( \frac{ x }{ x_\text{m} } \right)^\delta,
    \label{eq:paretodistribution}
\end{equation}
for all $x \geq x_\text{m}$ and is zero otherwise, where $x_\text{m} > 0$ is the cut-off or minimum value, and $\alpha > 0$ is the Pareto index. The Pareto distribution is of power law form, but has a finite integral and can therefore be normalised. The power law index $\delta < 0$ is the physically important scaling index of the cumulative FRB source count distribution and equals the Pareto index modulo the sign, $\delta = - \alpha$ We started from the Pareto maximum likelihood estimates $\hat{x}_\text{m} = \min_i (x_i)$ and $\hat{\alpha} = N / \sum_i \ln(x_i / \hat{x}_\text{m})$ \citep{1970Crawford, 2019James}, where $\min$ indicates the minimum and $N$ is the number of fluences. We fixed the cut-off value to $\hat{x}_\text{m}$, placed a mildly-informative Gaussian prior truncated at zero on the Pareto index ($\hat{\alpha}$ mean), and explored the posterior. Given the small FRB number regime that we are in, we verified the accuracy of our method on simulated data (see Appendix~\ref{sec:sourcecountverification}). Fig.~\ref{fig:sourcecounts} shows the measured integral fluence counts with the best-fitting Pareto distributions overlaid. The best-fitting power law exponents are $-1.9_{-0.7}^{+0.6}$ and $-0.8_{-0.3}^{+0.2}$ for the coherent and incoherent surveys, respectively. The CB discoveries are consistent with the Euclidean scaling, but the IB counts are significantly flatter (2.3-$\sigma$ significance). If we exclude the brightest FRB from the IB sample, the source count index steepens to $-1.2_{-0.5}^{+0.4}$, which is consistent with Euclidean within the errors.

In summary, the scaling of the FRB all-sky rates between surveys indicates a break or turn-over in the FRB fluence distribution below $\sim$1~Jy~ms. The MeerTRAP source counts within a survey are still uncertain due to the limited number of discoveries, but appear consistent with an Euclidean scaling.

Sophisticated joint analysis methods considering both primary FRB observables of S/N (fluence $F$) and DM, and eventually secondary distance information from the optical redshifts $z$ of secure host galaxy associations have recently been been developed and applied to the ASKAP, Parkes, and CHIME catalogue 1 samples \citep{2022James, 2022Shin}. They essentially modify the observed FRB rate in Eq.~\ref{eq:ratescaling} to a joint rate distribution $R_\text{FRB} (F, \text{DM}, z)$. Their analysis relies on having a sufficient number of FRBs per $F$, DM, and possibly $z$ bin. Applying such an analysis to the entire MeerTRAP FRB sample would certainly be a worthwhile exercise once appropriate FRB discovery numbers have been reached in the future.

\subsection{Constraints on the FRB repetition rate}
\label{sec:repetitionconstraints}

As part of the MeerTRAP survey, we spent approximately 27, 5, and 22~h in total on the three FRB discovery fields up to the end of 2021 (see Tab.~\ref{tab:burstproperties}). These were regular survey observations that the MeerTRAP instrument was commensal with. We inferred limits on the FRB repetition rates by assuming that the observable FRB sky rate above our detection threshold follows a Poisson distribution, i.e.\ neglecting any clustering in the burst arrival times that is reported for some repeaters, most notably FRB~20121102A \citep{2017Wang, 2018Oppermann, 2021Li}, and about which we have no a priori knowledge for these FRBs anyway. A memoryless Poisson process has a probability mass function given by
\begin{equation}
    P \left( k, \lambda \right) = \frac{ \lambda^k \: \exp (-\lambda) }{ k! },
    \label{eq:poisson}
\end{equation}
where $k$ is a natural number and $\lambda > 0$ is the Poisson parameter. It has a mean and expectation value of $\lambda$, which is related to the Poisson rate $R$, i.e.\ the number of events per unit time, by $\lambda = R t$. We estimated 95~per cent confidence level upper limits on the FRB repetition rate by using $\lambda_\text{max} (k = 1; p=0.95) = 4.744$ from \citet{1986Gehrels} to compute $R_\text{max} = \lambda_\text{max} / t_\text{obs}$, where $t_\text{obs}$ is the total exposure time on each FRB field from Tab.~\ref{tab:burstproperties}. The resulting upper limits are about 4.3, 23.4, and 5.2 bursts per day at the 95~per cent confidence level and above our detection threshold for FRBs~20201211A, 20210202D, and 20210408H, respectively. The total exposure primarily consisted of short pointings of $\sim$10~min duration for FRB~20210202D (pulsar timing) and somewhat longer ones $\sim$4.5~h for the other two FRBs (synthesis imaging). They were spaced quasi-regularly and semi-randomly in time due to the scheduling of the primary observing projects, over which MeerTRAP has no control. Our surveys are therefore sensitive to clustered burst arrivals and truly Poissonian behaviour, i.e.\ exponential waiting times.

The above treatment ignored the post-cursor burst detection of FRB~20210202D. If we consider it a genuine repeat pulse, its detection rate is $\sim$10$_{-8}^{+21}$ bursts per day at the 95~per cent Poisson confidence level and above our detection threshold. As above, this excludes any clustering effects that are likely at play. The rate is quite uncertain, as we only have about 4.9 hours of observing time on the discovery field of FRB~20210202D up to the end of 2021, the lowest exposure in the FRB sample presented here.

\subsection{Lack of band-limited FRBs}
\label{sec:bandlimitedbursts}

Aside from the data obtained in a dedicated follow-up campaign with MeerTRAP on the first repeater, FRB~20121102A \citep{2020Caleb, 2021Platts}, we did not discover any FRBs that show clear band-limited emission. In particular, all MeerTRAP FRBs published so far appear to have broadband emission across our $\sim$770~MHz of usable on-sky bandwidth at L-band. We did not find any credible candidates with spectral occupancies as low as seen, for example, in one of the repeat pulses at the Parkes \textit{Murriyang} telescope, i.e.\ a spectral width of only about 65~MHz \citep{2021Kumar}. This lack of band-limited FRBs suggests that our real-time transient search pipeline may be biased against them, especially as we only search the band-integrated data for performance reasons. However, we did indeed detect heavily-scintillated pulses from Galactic pulsars and RRATs, where we observed only a single narrow-band scintle within the band. Additionally, some MeerTRAP FRBs show characteristic scintillation patterns in their dynamic spectra \citep{2022Rajwade}, and we regularly detected pulses with emission restricted to the bottom part of the band. The latter are bursts from far out in the IB or CB response, where the high-frequency beam response is suppressed compared with those at lower frequencies (see Fig.~\ref{fig:beamarea}). This leads us to conclude that FRBs with narrow observed emission envelopes must be scarce in relation to those of at least $\sim$800~MHz width in the phase space probed by the MeerTRAP surveys. We estimated an upper limit for their all-sky rate using Eq.~\ref{eq:allskyrate}, the parameters of the MeerTRAP incoherent survey in Tab.~\ref{tab:allskyrates}, and a Poisson upper limit of 2.996 events at the 95~per cent confidence level given a non-detection \citep{1986Gehrels}. The all-sky rate of band-limited FRBs must be less than $890 \: \text{sky}^{-1} \text{d}^{-1}$, i.e.\ less than $\sim$40~per cent of the FRB all-sky rate inferred from the MeerTRAP incoherent survey above 3.44~Jy~ms.

This could have profound implications for the FRB population. If we assume that significantly band-limited bursts are primarily or only emitted by repeating FRBs, their number must be small compared with the whole population. This is consistent with the CHIME catalogue 1 sample, which suggested that only about four per cent of FRBs are repeaters \citep{2021CHIMECatalogue}. Additionally, strong scintillation of the order of 10 - 100~MHz bandwidth in the host galaxy or intervening ionised media, where only a single scintle falls within the recorded frequency range and the others are significantly suppressed, must be uncommon. With regards to narrow-band FRBs with higher spectral occupancy, broadband simple, narrow-band simple, and more complex bursts morphologies account for 30, 60, and 10 per cent of the CHIME catalogue 1 FRBs, respectively \citep{2021Pleunis}. That is, the majority are simple narrow-band bursts with typical bandwidths of $\sim$350-400~MHz for one-off events and $\sim$100-250~MHz for repeaters. While our estimate ($< 40$~per cent) could be compatible with the CHIME numbers within errors and accounting for the small sample size, the difference could indicate a genuine evolution of the observed FRB spectral occupancy with radio frequency or survey sensitivity (i.e.\ FRB population studied). In particular, it could be that the spectral occupancy decreases from L-band to CHIME frequencies (400-800~MHz) either intrinsically or due to propagation effects becoming more prominent.

\section{Discussion}
\label{sec:discussion}

\subsection{Is FRB~20210202D a repeater?}

The discovery of a faint post-cursor burst or emission component in FRB~20210202D is intriguing and makes it a good repeater candidate. Repeating FRBs often show the so-called ``sad trombone'' effect, i.e.\ complex time-frequency structure with subbursts that drift down in frequency with increasing time \citep{2019Hessels}. They also generally appear to have significantly wider burst profiles and are more band-limited than the apparent non-repeaters, at least at CHIME frequencies \citep{2021CHIMECatalogue}. FRB~20210202D exhibits none of those characteristics. However, repeaters also emit more broad-band spiky bursts, as seen for instance in FRB~20121102A \citep[``R1'';][]{2021Platts}, FRB~20180916B \citep[``R3'';][]{2020Marthi}, and FRB~20201124A \citep{2022Marthi}. Given its extremely narrow width, FRB~20210202D could be one of those spiky repeater bursts. Although narrower, it looks qualitatively similar to the broadband FRB~20221102A bursts with pre- or post-cursors presented in \citet{2021Platts}.

\subsection{Post-cursor burst separations}

FRB~20210202D is already the second MeerTRAP FRB in which a post-cursor burst was detected, with the other being FRB~20201123A \citep{2022Rajwade}. Interestingly, the observed post-cursor separations are surprisingly similar, with values of around 200~ms in each case. The FRBs are at the lower end of the DM distribution of the current MeerTRAP sample with observed DMs of $\sim$609 and $434~\text{pc}~\text{cm}^{-3}$ and extragalactic DMs of $\sim$486 and $109~\text{pc}~\text{cm}^{-3}$ above the Galactic ISM and halo contributions. This could indicate that they are indeed reasonably nearby repeaters, especially FRB~20201123A. Aside from this, their parameters differ significantly. For instance, FRB~20201123A's pulse width is about double that of FRB~20210202D's. In comparison, the histogram of the sub-burst separations in the CHIME catalogue 1 sample peaks around $\sim$10~ms with only two bursts above 30~ms and a maximum  separation near 65~ms. Out of those, repeaters seems to show somewhat larger values \citep{2021Pleunis}. The $\sim$200~ms separations in the two MeerTRAP FRBs is significantly larger than this, which supports the idea that they are faint repeat pulses.

Why are their post-cursor separations almost precisely the same? We are not aware of any obvious instrumental reasons for why that should be the case. This has neither been seen so far in other MeerTRAP FRBs nor any of thousands of pulsar or RRAT pulses. A shift in arrival time of one or multiple frequency sub-bands could happen in exceptional cases when the beam-former nodes get out of sync. However, the sky signal would get shifted in time and not copied. We are currently commissioning a real-time system to write out voltage data whenever FRBs are discovered. This will allow us in the future to test whether similar post-cursors are coherent copies of the primary bursts and determine their polarisation properties. It is hard to imagine how a delayed mirror image of the primary burst could be introduced into the data stream, and we conclude that it must be astrophysical. We also caution that these are very small number statistics. Nonetheless, perhaps the 200~ms separation corresponds to an oscillation frequency, activity or rotation period, or any of its harmonics in the FRB progenitor or its emission mechanism. Or maybe it is related to the quasi-periodic sub-components that have been reported in some FRBs \citep{2022CHIMEPulsePeriodicity, 2022PastorMarazuela}. Most notable here is FRB~20191221A with a closely comparable and statistically significant component periodicity of 216.8~ms \citep{2022CHIMEPulsePeriodicity}. Similarly, it could be a significantly scaled-up version of the quasi-periodic microstructure observed in several radio pulsars \citep{1990Cordes}. Alternatively, the post-cursor burst might be an attenuated echo of the primary, for which the separation would correspond to a light travel time difference. The same is true in the case of gravitational lensing of FRBs. For example, our data captures around the bursts are sensitive to FRB millilensing with delays of $\sim$milliseconds and above by intermediate-mass black holes or dark matter halos \citep{2022Connor}. The lensed copies of the FRB signal will be fainter than the primary burst, as in our post-cursors. The phenomenon offers exciting prospects for studying cosmology and fundamental physics using FRBs \citep{2014Zheng, 2018LiLensing}. Our new voltage buffer dump system will allow us to test if that is the case too.

The inter-burst arrival or waiting times between bursts from repeaters are of scientific interest and have been studied by several authors. Already early on it was realised that their bursts arrive often clustered in time \citep{2017Wang, 2018Oppermann}. For instance, both FRBs~20121102A and 20200120E show clustering seen as bimodality in their waiting time distributions. The short-duration clustering is most relevant for this discussion. In FRB 20121102A, the fast clustering peak occurs around 22 to 24~ms \citep{2022Hewitt, 2023Jahns} if sub-bursts are excluded and around 3.4~ms if they are not \citep{2021Li}. In FRB 20200120E, the fast peak in the waiting time distribution appears around 1~s \citep{2022Nimmo}. Neither values are close to the $\sim$200~ms separation seen here and they differ already significantly among the two repeaters. Hence, it is unclear whether the post-cursors are sub-bursts or repeat pulses. A larger sample of well-constrained repeater waiting time distributions is needed to inform the distinction.

\subsection{A deficit of low-fluence FRBs}

The FRB all-sky rate inferred from the MeerTRAP coherent survey is significantly below that expected from the best-fitting power law scaling from surveys at higher limiting fluences $\geq 1$~Jy~ms, see Fig.~\ref{fig:allskyrate}. Equivalently, the power law scaling between the MeerTRAP coherent and incoherent surveys is appreciably flatter than that among the high-fluence surveys. The flatter power law scaling from MeerTRAP extrapolates near the FAST rate at a limiting fluence $\sim$45 times lower. The flattening of the scaling of the FRB all-sky rate with limiting fluence and the apparent deficit of low-fluence FRBs could have important implications for the FRB population and cosmology.

In the following, we discuss several possible explanations for the FRB deficit. (1) The MeerTRAP coherent rate is only based on 6 CB detections and, therefore, still in the small number statistics regime. Further detections might either strengthen the trend or reduce the tension with the high-fluence estimates. (2) The rate inferred from the MeerTRAP coherent survey could be slightly underestimated due to the more complex FoV than that of the incoherent survey, perhaps even by a factor of two. However, it is unlikely to be off by an order of magnitude. To illustrate the point, when extrapolating from the best-fitting high-fluence power law down to the MeerTRAP coherent survey fluence limit (shown as a slightly transparent yellow line in Fig.~\ref{fig:allskyrate}), we would expect to detect a rate of $25_{-12}^{+21} \times 10^3$ instead of $8.2_{-4.6}^{+8.0} \times 10^{3} \: \text{sky}^{-1} \: \text{d}^{-1}$ in the MeerTRAP coherent survey, which translates to about $34_{-16}^{+28}$ FRB CB detections above the completeness threshold instead of 6. Where are those missing FRBs? It seems unlikely that we missed such a large number of FRBs in our detection pipeline. (3) The MeerTRAP fluence completeness threshold estimates could be systematically off. If both survey completeness limits were shifted by the same amount, the power law exponent would be preserved. Shifting the CB survey fluence limit up would still mismatch the absolute rate expected from the high-fluence scaling. However, it would steepen the MeerTRAP intra-survey power law exponent closer to the high-fluence value. (4) The lack of low-fluence FRBs could naturally be explained by a genuine break or turn-over in the rate -- fluence threshold relation below 2~Jy~ms. It could, for example, indicate that FRBs transition from the Euclidean scaling ($\propto F_c^{-1.5}$) to the constant scaling ($\propto F_c^{-1}$) in that fluence range. Astrophysically, this could be due to the FRB population's cosmic evolution in redshift or luminosity space or the Universe's expansion, which both flatten the FRB fluence distribution \citep{2018Macquart}. Those effects would only become important for higher-redshift FRBs. Our analysis is consistent with that of \citet{2019James} who hinted at the existence of a low-fluence downturn or equivalently a high-fluence steepening based on an early sample of ASKAP and Parkes FRBs. (5) More simplistically, the more sensitive surveys might detect more distant populations of FRBs, which have shallower fluence distributions. We can see that when comparing the median FRB DMs of various surveys arranged from shallow to deep. The ASKAP sample has a median DM of $\sim$400~$\text{pc}~\text{cm}^{-3}$ \citep{2018Shannon}, the CHIME sample a median DM of $\sim$500~$\text{pc}~\text{cm}^{-3}$ \citep{2021CHIMECatalogue}, the Parkes sample a median DM of $\sim$900~$\text{pc}~\text{cm}^{-3}$ \citep{2018Shannon}, and the entire MeerTRAP sample considered here has a median DM of $\sim$740~$\text{pc}~\text{cm}^{-3}$.

Irrespective of the origin of the discrepancy, it will be interesting to see whether future MeerTRAP CB discoveries and improved beam or survey modelling reduce the tension to the high-fluence results.

\section{Conclusions}
\label{sec:conclusions}

In this paper, we presented a sample of three well-localised FRBs discovered with the newly-commissioned MeerTRAP transient search instrument at the MeerKAT telescope array in South Africa. We analysed their burst properties and showed their localisations within a multi-wavelength context. We conclude the following.

Each FRB was discovered in the data from a single coherent tied-array beam. Based on the non-detections in adjacent beams, we localised them to about $1~\text{arcmin}^2$ or better. Therefore, they are more precisely localised than about 97~per cent of the currently published FRBs.

All the FRBs occurred in the southern hemisphere, at high absolute Galactic latitudes over $\sim$30~deg.

They have substantial observed DMs between about 609 and $1196~\text{pc}~\text{cm}^{-3}$, with extragalactic contributions between about 490 and $1100~\text{pc}~\text{cm}^{-3}$, indicating expected host galaxy redshifts from as low as 0.2 up to about 1.2.

The FRBs have refined S/N values of at least $\sim$15, meaning they are robust detections. On the other hand, their inferred fluences of $> 0.4$~Jy~ms place them at the low-fluence end of the known FRB population.

We tried to associate the FRBs to host galaxy candidates from the literature. Our analyses are mostly inconclusive, as several galaxies within the localisation regions have non-negligible association probabilities. The exception is FRB~20210408H, for which there are only four host galaxy candidates. We derived a photometric redshift of $z_\text{phot} = 0.45 \pm 0.08$ for the favoured host ($p(O|x) \simeq 0.35 - 0.53$), galaxy 1 (PS1 ID 74052043311949899). While lower than expected, the redshift is compatible with the FRB's DM of almost $1196~\text{pc}~\text{cm}^{-3}$ at the 2-$\sigma$ level when assuming a moderate host DM contribution $\geq 150~\text{pc}~\text{cm}^{-3}$ and taking into account the uncertainty in the DM -- redshift relation. Alternatively, the galaxy might be an unrelated foreground galaxy, and the actual host is not visible in our current imaging data. The probability of an unseen host is 34 per cent.

The FRBs are mostly unresolved in our data due to the broad channelisation and the effects of intra-channel dispersive smearing. FRB~20201211A exhibits hints of a marginally significant scattering contribution at the 1 to 2-$\sigma$ level.

FRB~20210202D appears to be followed by a faint post-cursor pulse about 200~ms after the main burst component. The FRB is a good repeater candidate, although it does not show any typical repeater-like characteristics. We speculated that it is a broad-band spiky repeater burst.

Additionally, we analysed the properties of the two simultaneous MeerTRAP transient surveys at L-band based on the entire sample of 11 FRBs discovered by the end of 2021.

We used conventional approaches to estimate fluence completeness thresholds of 0.66 and 3.44~Jy~ms for the coherent and incoherent MeerTRAP surveys, respectively.

Between 2019 June and the end of 2021 December, the MeerTRAP instrument spent approximately 317.5~d surveying the sky. Excluding known periods of reduced pipeline performance, and based on the entire FRB sample discovered in that time, we inferred FRB all-sky rates of $8.2_{-4.6}^{+8.0}$, $2.1_{-1.1}^{+1.8}$, and $1.7_{-1.0}^{+1.8} \times 10^3 \: \text{sky}^{-1} \: \text{d}^{-1}$ at 1.28~GHz above 0.66, 3.44, and 3.44~Jy~ms and assuming 50~per cent detection efficiency.

The power law scaling between the MeerTRAP FRB all-sky rates is flatter than those in the literature obtained at higher limiting fluences $\geq 1$~Jy~ms at the 1.4-$\sigma$ confidence level. There appears to be a shortage of low-fluence FRBs, suggesting a break or turn-over in the rate versus fluence relation below 2~Jy~ms. We speculated that the deficit could be progenitor-intrinsic or due to cosmological effects. Perhaps we see signs of progenitor evolution. The MeerTRAP coherent survey is one of the first to systematically explore the FRB population's low-fluence end. Although the numbers of our current FRB discoveries are limited, the CB cumulative source count distribution within the survey appear to follow the Euclidean $\propto F^{-3/2}$ scaling. The IB counts are significantly flatter, but become consistent with Euclidean if the brightest IB FRB is excluded.

We constrained the repetition rates of the three FRBs to less than 4.3, 23.4, and 5.2 bursts per day at the 95~per cent confidence level. If we include FRB~20210202D's post-cursor as a genuine repeat pulse, its detection rate is $\sim$10$_{-8}^{+21}$ bursts per day above our detection threshold at the 95~per cent Poisson confidence level.

No clear band-limited FRBs were discovered. This suggests that they are scarce for our observing setup compared with FRBs with more band-filling emission. Their inferred all-sky rate must be less than $890 \: \text{sky}^{-1} \text{d}^{-1}$, i.e.\ less than about 40~per cent of the incoherent survey rate above a limiting fluence of 3.44~Jy~ms.

\section*{Acknowledgements}

We thank the reviewer for their insightful comments and questions that significantly improved the paper. The authors thank the MeerKAT Large Survey Project teams for allowing MeerTRAP to observe commensally. FJ, MCB, MC, LND, MM, VM, KMR, SS, BWS, and MPS acknowledge funding from the European Research Council (ERC) under the European Union's Horizon 2020 research and innovation programme (grant agreement no.~694745). We also acknowledge the use of TRAPUM infrastructure funded and installed by the Max-Planck-Institut f{\"u}r Radioastronomie and the Max-Planck-Gesellschaft. We thank Chris Williams for his assistance in getting the MeerTRAP pipeline up and running. MC acknowledges the support of an Australian Research Council Discovery Early Career Research Award (project number DE220100819) funded by the Australian Government and the Australian Research Council Centre of Excellence for All Sky Astrophysics in 3 Dimensions (ASTRO 3D) through project number CE170100013. KMR acknowledges support from the Vici research programme ``ARGO'' with project number 639.043.815, financed by the Dutch Research Council (NWO). JXP as a member of the Fast and Fortunate for FRB Follow-up team, acknowledges support from NSF grants AST-1911140 and AST-1910471. The MeerKAT telescope is operated by the South African Radio Astronomy Observatory, which is a facility of the National Research Foundation, an agency of the Department of Science and Innovation.

\section*{Data availability}

The localisation region files, \textsc{healpix} localisation maps, and the meta data for the three MeerTRAP FRB discoveries presented in this paper are available from our Zenodo repository at \url{https://doi.org/10.5281/zenodo.6047539}. Other data underlying this article will be shared upon reasonable request to the corresponding author.





\bibliographystyle{mnras}
\bibliography{frb_paper}



\appendix


\section{Verification of our FRB source count estimation method}
\label{sec:sourcecountverification}

We tested how accurately our analysis method estimated the scaling indices of the cumulative or integral source count distributions with emphasis on the low number regime. We did that by randomly drawing synthetic FRB fluence data from Pareto distributions with parameters close to those of our data sets. Specifically, we used values of (0.66, 1.9) and (3.44, 0.8) for the $x_\text{m}$ and $\alpha$ parameters in Eq.~\ref{eq:paretodistribution} for the simulated CB and IB fluences. These Pareto indices $\alpha$ correspond to power law exponents $\delta$ of $-1.9$ and $-0.8$, respectively. We successively drew 50, 10, and 6 random samples from each distribution with an equal number for each simulated survey and ran those synthetic data through our estimation software. We repeated the process 240 to 330 times to check the spread in returned measurements. For 50 FRBs, the recovered values are well within the 1-$\sigma$ fit errors from the injected ones, with a sample spread of 0.1 (IB) and 0.3 (CB) standard deviations. For 10 FRBs, the sample variation becomes more significant as the probability of missing the rare high-fluence events increases, especially for the steeper CB distribution. The median recovered $\alpha$ values are steeper (higher) than the injected parameters by 0.2 and 0.3. The sample standard deviations are 0.4 and 0.8, i.e.\ there is a significant scatter towards steeper indices in the CB sample. When bright bursts are present, the recovered indices match the injected values within the fit uncertainties. For 6 FRBs, the fit errors are appreciably larger than before. In most cases, the estimated indices are compatible with the injected values within the 1-$\sigma$ uncertainties. The sample medians are steeper by 0.3 (IB) and 0.7 (CB), and the sample standard deviations amount to 0.9 and 1.4, respectively. In summary, detecting the rare bright FRBs is crucial to accurately characterise the population's fluence distribution. Without them, the measurements are biased to exponents that are too steep with respect to the true underlying distribution.


\bsp	
\label{lastpage}
\end{document}